\documentclass[aps,prd,superscriptaddress,11pt,showpacs,notitlepage]{revtex4}
	\usepackage{graphicx}
	\usepackage{amsmath,amssymb,mathrsfs}
	\usepackage[colorlinks=true, a4paper=true, pdfstartview=FitV,
linkcolor=blue, citecolor=blue, urlcolor=blue]{hyperref}

\usepackage{epsf}
\usepackage{epsfig}
\usepackage[usenames,dvipsnames]{xcolor}

\newcommand{\be}{\begin{equation}}
\newcommand{\ee}{\end{equation}}
\newcommand{\bea}{\begin{eqnarray}}
\newcommand{\eea}{\end{eqnarray}}
\newcommand{\pa}{\partial}
\newcommand{\bb}{\bibitem}
 
\newcommand{\eqn}{\begin{eqnarray}}
\newcommand{\eqnx}{\end{eqnarray}}
\newcommand{\bn}{\mbox{\boldmath $n$}}
\newcommand{\bx}{\mbox{\boldmath $x$}}

\numberwithin{equation}{section}

\begin{document}

\title{The BPS sectors of the Skyrme model and their non-BPS extensions}
\author{C. Adam}
\affiliation{Departamento de F\'isica de Part\'iculas, Universidad de Santiago de Compostela and Instituto Galego de F\'isica de Altas Enerxias (IGFAE) E-15782 Santiago de Compostela, Spain}
\author{D. Foster}
\affiliation{School of Physics, HH Wills Physics Laboratory, University of Bristol, Tyndall Avenue, Bristol BS8 1TL, United Kingdom}
\author{S. Krusch}
\affiliation{School of Mathematics, Statistics and Actuarial Science, University of Kent, Canterbury CT2 7FS, United Kingdom}
\author{A. Wereszczynski}
\affiliation{Institute of Physics,  Jagiellonian University,
Lojasiewicza 11, Krak\'{o}w, Poland}

\begin{abstract}
Two recently found coupled BPS submodels of the Skyrme model are further analyzed. Firstly, we provide a geometrical formulation of the submodels in terms of the eigenvalues of the strain tensor. 
Secondly, we study their thermodynamical properties and show that the mean-field equations of state coincide at high pressure 
and read $p=\bar{\rho}/3$. We also provide evidence that matter described by the first BPS submodel has some similarity with a Bose-Einstein condensate. 
Moreover, we show that extending the second submodel to a non-BPS model by including certain additional terms of the full Skyrme model does not spoil the respective ansatz, leading to an ordinary differential equation for the profile of the Skymion, for any value of the topological charge. This allows for an almost analytical description of the properties of Skyrmions in this model. In particular, we analytically study the breaking and restoration of the BPS property.
Finally, we provide an explanation of the success of the rational map ansatz.
\end{abstract}


\maketitle 
%%%%%%%%%%%%%%%%%%%%%%%%%%%%%%%%%%%%%%%%%
\section{Introduction}
%%%%%%%%%%%%%%%%%%%%%%%%%%%%%%%%%%%%%%%%%
The Skyrme model \cite{skyrme} is a highly nonlinear effective model of nuclear physics, which has had some success in replicating nuclear states. A consequence of this nonlinearity is that physically interesting solutions, called Skyrmions, have only been identified numerically revealing very sophisticated geometrical structures \cite{L24}, \cite{L024}. The task of finding soliton solutions in Skyrme-like models is sometimes simplified by the possibility to reduce their second order Euler-Lagrange equations to first order equations - the so-called BPS equations. The corresponding BPS solutions then saturate a topological energy bound known as the Bogomolny bound. 
The Skyrme model \cite{skyrme} is not BPS, and hence its equations of motion cannot be reduced to first order equations.
However, it is possible to identify certain {\em BPS submodels} of the Skyrme model \cite{BPS}, \cite{newBPS}. The understanding and analysis of these BPS submodels is important for several reasons. 

First of all, the much simpler solutions of the BPS submodels may reveal certain qualitative properties of full Skyrmions in an analytically tractable way, or they may provide useful starting points for a numerical treatment of the full model. 

A second reason is related to another significant problem in applying the Skyrme model to nuclear physics, namely that its static Skyrmion solutions are too tightly bound to replicate experimentally observed nuclei. Conversely, BPS Skyrmions have zero classical binding energies by construction. Consequently,
there have been proposals to solve this problem 
by the inclusion of the so-called BPS Skyrme model \cite{BPS}, \cite{prl}, \cite{mar} consisting of the sextic and potential term or by inclusion of infinitely many vector mesons \cite{vecSut}, \cite{rho}. Alternatively, one can add suitably chosen repulsive potentials to reduce binding energies \cite{Leeds}, \cite{Bjarke}. Furthermore, vibrational quantisation of Skyrmions can reduce binding energies \cite{vib}.

Quite recently, BPS submodels of the original Skyrme model \cite{skyrme} have been identified \cite{newBPS}. This allows, within each submodel, to reduce the static field equations to more tractable first order differential equations. The two submodels constitute the original Skyrme model in the sense that the full model is a sum of these two submodels, but they are not proper Skyrme models on their own.  This means that eliminating one submodel by a suitable choice of coupling constants, simultaneously eliminates the other one. As they always appear together, it is natural to call them {\it coupled} BPS submodels. Nonetheless, one can study each submodel {\it separately}, and their solutions reveal interesting properties of Skyrmions. For example, it has been shown that the origin of the success of the rational map ansatz approximation has its roots in the first coupled BPS submodel which has rational maps as {\it exact} solutions. 

The existence of such coupled sectors give us a unique possibility for some analytical insight into the complicated structure of the full theory. Our aim is to further understand the properties of the coupled BPS submodels and of certain extensions which result from the inclusion of additional Skyrme model terms.

\vspace*{0.2cm}

The Skyrme model \cite{skyrme} can be expressed as 
\begin{align}
\mathcal{L}&=c_2\mathcal{L}_2+c_4\mathcal{L}_4+c_6\mathcal{L}_6+c_0\mathcal{L}_0 ~~\mbox{with} \\
\mathcal{L}_{2} =-\frac{1}{2} &\mbox{ Tr } L_\mu L^\mu,~\mathcal{L}_4=  \frac{1}{16} \mbox{ Tr } [L_\mu, L_\nu ]^2,~ {\rm and} ~ \mathcal{L}_6=-\pi^4  \mathcal{B}_\mu \mathcal{B}^\mu,  \nonumber
\end{align}
where $L_\mu = U^\dag \pa_\mu U$ with $U\in \mbox{SU(2)},$ and $c_i$ are dimensionful, non-negative coupling constants. 
This has an associated baryon current, $\mathcal{B}^\mu = \frac{1}{24\pi^2} \epsilon^{\mu \nu \rho \sigma} \mbox{Tr} \; L_\nu L_\rho L_\sigma$, and an invariant baryon number $B=\int \mathcal{B}^0 d^2x\in\mathbb{Z}$. Further, $\mathcal{U} = -\, \mathcal{L}_0$ is a potential. In most of what follows, we will assume $c_2 =1$, $c_4 =1$, which may always be achieved by an appropriate choice of units of length and energy.

The conventional Skyrme model, $\mathcal{L}_{24}=\mathcal{L}_2+\mathcal{L}_4$, is not a BPS type model. Recently, it has been shown in \cite{newBPS} that when the Skyrme field is re-expressed as
\be
\label{comp-cord}
U
=\exp(i\xi(\bx) \tau\cdot \bn(\bx)) 
\quad {\rm with} \quad
\bn(\bx)
=\frac{1}{1+|u(\bx)|^2}\left(
\begin{array}{c}
u(\bx)+\bar{u}(\bx)\\-i(u(\bx)-\bar{u}(\bx))\\1-|u(\bx)|^2
\end{array}
\right), 
\ee
the Skyrme model $\mathcal{L}_{24}$ can be viewed as a sum
 of two coupled BPS submodels
\bea
 \mathcal{L}_{24} = \mathcal{L}_{24}^{(1)} +  \mathcal{L}_{24}^{(2)} , \nonumber 
 \eea  
where $\mathcal{L}_{24}^{(1)}$ and $\mathcal{L}_{24}^{(2)}$ are the two BPS submodels
\be \label{L(1)}
\mathcal{L}_{24}^{(1)} \;\; = \; \; 4\sin^2\xi \frac{u_\mu \bar{u}^\mu}{(1+|u|^2)^2}  -  4 \sin^2 \xi \left( \xi_\mu \xi^\mu \frac{u_\nu \bar{u}^\nu}{(1+|u|^2)^2} -\frac{\xi_\mu \bar{u}^\mu \;\xi_\nu u^\nu}{(1+|u|^2)^2} \right)
\ee
and
\be
  \mathcal{L}_{24}^{(2)} \;\; = \; \;  \xi_\mu \xi^\mu  -    4\sin^4 \xi  \frac{(u_\mu \bar{u}^\mu)^2-u_\mu^2 \bar{u}^2_\nu }{(1+|u|^2)^4},
\ee
where $\xi_\mu \equiv \partial_\mu \xi$ and $u_\mu \equiv \partial_\mu u.$ 

The first BPS submodel $\mathcal{L}_{24}^{(1)}$  gives rise to the Bogomolny equation
\be \label{(1)-bog}
u_i\pm i \epsilon_{ijk} \xi_j u_k=0,
\ee
and its complex conjugate. These equations imply the constraints
\be \label{constr-1}
u_i \xi_i=\bar{u}_i \xi_i=0 \quad {\rm and} \quad u_i^2=\bar{u}_j^2=0 .
\ee
The second BPS submodel $\mathcal{L}_{24}^{(2)}$ leads to the Bogomolny equation,
\be \label{(2)-bog}
\xi_i \mp  \frac{2i\sin^2 \xi}{(1+|u|)^2} \epsilon_{ijk} u_j \bar{u}_k =0,
\ee
 implying the constraints,
\be \label{constr-2}
u_i\xi_i = \bar{u}_i \xi_i=0.
\ee
These two BPS models independently have the topological bounds
\be
E^{(1)} \geq 8\pi^2 |B| \quad {\rm and} \quad E^{(2)} \geq 4\pi^2 |B|, 
\ee
where $E^{(1)}$ and $E^{(2)}$ are the energies of the first and second BPS submodels, respectively. 

There is a third unrelated BPS submodel, the so-called BPS Skyrme model, where conventionally $c_0 = m^2$ and $c_6 = \lambda^2$,
\be
\label{LBPS}
\mathcal{L}_{BPS}= \lambda^2 \mathcal{L}_6+m^2 \mathcal{L}_0,
\ee
which has the BPS static field equations
\be \label{BPS-bog}
\lambda \frac{ \sin^2 \xi}{(1+|u|^2)^2} \; i \epsilon^{ijk} \xi_{i} u_{j} \bar{u}_{k} =\pm   m \sqrt{\mathcal{U}} .
\ee

Note that the BPS Skyrme model is a proper submodel, since it can be found as a certain limit in the 4-dimensional parameter space of the full theory while, as we pointed out before, the first and second coupled BPS submodels do not have this property. The coupled submodels always exist together because each of them receives contributions both from the quadratic Dirichlet and from the quartic Skyrme  term. In spite of this fact, it is interesting to study both models separately, both because of their simplicity and because they reveal crucial mathematical and physical features of Skyrmions in the full model.  Of course, if treated separately they do not cover the whole variety of phenomena in the Skyrme model $\mathcal{L}_{24}.$ 

In the present work, we want to further analyze the two coupled BPS submodels, especially from the thermodynamical point of view (type of matter, mean-field equations of state). We also analytically investigate the solutions of the new BPS submodels once new terms, such as a potential or the sextic term, are added. In particular, we are interested in deformations of the BPS submodels which preserve the ansatz for the $\vec{n}\in \mathbb{S}^2$ part of the Skyrme field, thus still allowing for the almost analytical calculation of solutions for any value of the topological charge.
%%%%%%%%%%%%%%%%%%%%%%%%%%%%%
\section{Geometric meaning}
%%%%%%%%%%%%%%%%%%%%%%%%%%%%%
In order to develop some geometric understanding of these new coupled BPS submodels and their Bogomolny equations, we use the well know formulation of the static energy integral of the $\mathcal{L}_{24}$ Skyrme model in terms of the eigenvalues $\lambda_i^2$ of the strain tensor \cite{sphere}, \cite{SM},
\be
D_{ij}=-\frac{1}{2} \mbox{ Tr } (L_iL_j).
\ee
Using equation \eqref{comp-cord} the three eigenvalues of the strain tensor $\lambda_1^2,\lambda_2^2,\lambda_3^2$ can be expressed as,
\be
\lambda_1 \lambda_2 \lambda_3 = \pm 2\frac{\sin^2 \xi}{(1+|u|^2 )^2} (i\epsilon^{ijk}\xi_i u_j \bar u_k ),
\ee
\be
\lambda_1^2 + \lambda_2^2 + \lambda_3^2 = \xi_i^2 + 4 \sin^2 \xi \frac{u_i \bar u^i}{(1+|u|^2)^2},
\ee
and
\be 
\lambda_1^2 \lambda_2^2 + \lambda_1^2 \lambda_3^2 + \lambda_2^2 \lambda_3^2 = 
4\sin^4 \xi \frac{(u_i \bar u^i)^2 - u_i^2 \bar u_j^2 }{(1+|u|^2)^4} + 
   4\sin^2 \xi \left( \xi_i^2 \frac{u_j\bar u^j}{(1+|u|^2)^2} - \frac{\xi_iu^i \xi_j \bar u^j}{(1+|u|^2)^2} \right) .
\ee
Hence,
\be
E_{24}=\int d^3 x (\lambda_1^2+\lambda_2^2+\lambda_3^2 +  \lambda_1^2\lambda_2^2+\lambda_2^2\lambda_3^2+\lambda_2^2\lambda_1^3) .
\ee
The derivation of the topological Skyrme-Faddeev bound is now straightforward 
\bea
E&=& \int d^3 x \left( (\lambda_1 \pm \lambda_2 \lambda_3)^2+(\lambda_2 \pm \lambda_3 \lambda_1)^2+(\lambda_3 \pm \lambda_1 \lambda_2)^2  \right) \mp 6 \int d^3 x  \lambda_1 \lambda_2 \lambda_3, \nonumber \\
&\geq& 6  \left| \int d^3 x  \lambda_1 \lambda_2 \lambda_3 \right| = 12\pi^2 B,
\eea
where $B$ is the baryon charge associated with the baryon 
density
\be
\label{Bden}
\mathcal{B}=\frac{1}{2\pi^2}  \lambda_1 \lambda_2 \lambda_3 .
\ee
This topological bound can be saturated if and only if the following Bogomolny equations are satisfied
\be
\lambda_1 = \pm \lambda_2 \lambda_3, \;\; \lambda_2 = \pm \lambda_3 \lambda_1, \;\; \lambda_3 = \pm \lambda_1 \lambda_2 .
\label{BPS-L}
\ee
However, it is known that there are no non-trivial solutions satisfying these equations on $\mathbb{R}^3$. The case $\mathbb{S}^3$ is discussed in the next section.
 
Let us now analyse these Bogomolny equations in the separation of variables ansatz, where we chose the plus sign.
Furthermore we decompose $u=ge^{i\Phi}$, set $G=g^2$ and use the invariant gradient notation 
\be
\nabla = \hat e_x \partial_x + \hat e_y \partial_y + \hat e_z \partial_z = \hat e_r \partial_r + \frac{1}{r} \hat e_\theta \partial_\theta + \frac{1}{r\sin\theta} \hat e_\varphi \partial_\varphi, 
\ee
to obtain
\be
\lambda_1 \lambda_2 \lambda_3 = \pm 2\frac{\sin^2 \xi}{(1+G)^2} \nabla \xi \cdot (\nabla G \times \nabla \Phi ),
\ee
\be
\lambda_1^2 + \lambda_2^2 + \lambda_3^2 = (\nabla \xi)^2 + 4\frac{\sin^2 \xi}{(1+G)^2}\left( \frac{1}{4G}(\nabla G)^2 + G (\nabla \Phi)^2 \right),
\ee
\bea
\lambda_1^2 \lambda_2^2 + \lambda_1^2 \lambda_3^2 + \lambda_2^2 \lambda_3^2 &=& 
\frac{4\sin^2 \xi}{(1+G)^2} \left( \frac{\sin^2 \xi}{(1+G)^2} \left( (\nabla G)^2 (\nabla \Phi)^2 - (\nabla G\cdot \nabla \Phi )^2 \right) +
\right. \nonumber \\
&& \left.
\hspace*{-1.0cm} \frac{1}{4G} \left( (\nabla \xi)^2 (\nabla G)^2 - (\nabla \xi \cdot \nabla G)^2 \right) + G\left( (\nabla \xi)^2 (\nabla \Phi)^2 - (\nabla \xi \cdot \nabla \Phi)^2 \right) \right).
\eea
Now we introduce spherical polar coordinates and assume $\xi = \xi (r)$ and $u=u(\theta ,\varphi)$. This is consistent with constraint \eqref{constr-2} which is satisfied by both BPS submodels and can be written as 
\be
\label{const-2a}
\nabla \xi \cdot \nabla u = 0 \quad {\rm and} \quad \nabla \xi \cdot \nabla {\bar u} = 0.
\ee
In fact it is sufficient to assume $\xi = \xi(r)$ and then $u=u(\theta,\varphi)$ follows from constraint \eqref{const-2a}.
This assumption implies that the strain tensor in spherical polar coordinates partially diagonalises such that one eigenvalue (let us say, $\lambda_1^2$) is equal to $\xi_r^2$,  $\lambda_1^2 = \xi_r^2$. We choose $\lambda_1 = \xi_r $ and the plus sign (or $\lambda_1 = -\xi_r$ and the minus sign), leading to
\bea
\lambda_2\lambda_3&=&2\frac{\sin^2 \xi}{(1+ |u|^2)^2} i \hat e_r \cdot (\nabla u \times \nabla \bar u), \nonumber
\\
\lambda_2^2 + \lambda_3^2 &=& 4\sin^2 \xi \frac{\nabla u \cdot \nabla \bar u}{(1+|u|^2)^2}, \nonumber
\\
\lambda_2^2 \lambda_3^2 &=& 4\sin^4 \xi \frac{(\nabla u \cdot \nabla \bar u)^2 - (\nabla u)^2 (\nabla \bar u)^2}{(1+|u|^2)^4} .
\eea
The third equation is a consequence of the first, but it is nevertheless useful to see directly the consequence of the complex eikonal equation $(\nabla u)^2 =0$. On the one hand, the complex eikonal equation implies that $u(\theta ,\varphi)$ is either a holomorphic or an antiholomorphic function, so that $u$ can be written as $u=u(z)$ or $u=u(\bar z)$ where $z= \tan \frac{\theta}{2} e^{i\varphi}$. On the other hand, the complex eikonal equation immediately implies that 
\be
\lambda_2^2 = \lambda_3^2 = 2\sin^2 \xi \frac{\nabla u \cdot \nabla \bar u}{(1+|u|^2)^2}.
\ee
For the second type of BPS system, the expressions in terms of $\xi$, $G$ and $\Phi$ are more useful. Indeed, using the separation of variables ansatz $\xi = \xi (r)$, $G=G(\theta)$, $\Phi = \Phi (\varphi)$, the equations for the eigenvalues simplify to
\bea
\lambda_1^2 &=& (\nabla \xi)^2 = \xi_r^2, 
\\
\lambda_2^2 &=& \frac{\sin^2 \xi}{G(1+G)^2}(\nabla G)^2 = \frac{\sin^2 \xi \, G_\theta^2 }{G(1+G)^2 r^2}, 
\\
\lambda_3^2 &=& 4\frac{\sin^2 \xi}{(1+G)^2}G(\nabla \Phi)^2 = 4\frac{\sin^2 \xi \, G \, \Phi_\varphi^2}{(1+G)^2 r^2 \sin^2 \theta } .
\eea
Now we want to relate the BPS equations for the two submodels to equations for the eigenvalues $\lambda_i$.
Multiplying the BPS equation (\ref{(2)-bog}) of the second submodel $\mathcal{L}_{24}^{(2)}$ by $\hat e_r$ (multiplication by the other two basis vectors perpendicular to $\hat e_r$ gives zero), we obtain the equation
\be
\xi_r =    \pm 2\frac{\sin^2 \xi}{(1+ |u|^2)^2} i \hat e_r \cdot (\nabla u \times \nabla \bar u) \quad \Leftrightarrow 
\quad \lambda_1 = \pm \lambda_2 \lambda_3 .
\ee
The first BPS equation (\ref{(1)-bog}) reads $\nabla u = \mp i \nabla \xi \times \nabla u$ and implies the constraints $(\nabla u)^2 =0$ and $\nabla \xi \cdot \nabla u =0$. 
Multiplying the BPS equation by $\nabla \bar u$ results in 
\be
\nabla u \cdot \nabla \bar u = \pm \xi_r i \hat e_r \cdot (\nabla u \times \nabla \bar u) \quad \Leftrightarrow \quad \frac{1}{2} \left( \lambda_2^2 + \lambda_3^2 \right) = \mp \lambda_1 \lambda_2 \lambda_3.
\ee
But the constraint $(\nabla u)^2 =0$ for this ansatz implies that $\lambda_2^2 = \lambda_3^2$ which directly leads to $\lambda_1 = \xi_r = \pm 1$, which is the radial BPS equation for this submodel, see \cite{newBPS}. 

For both submodels we find that, after a separation of variables ansatz $\xi = \xi (r)$, $u=u(\theta ,\varphi)$, their BPS equations may be expressed as simple algebraic equations for the eigenvalues of the strain tensor. The BPS equation for the second submodel is equivalent to $\lambda_1 = \pm \lambda_2 \lambda_3$, whereas the BPS equation of the first model is equivalent to the two equations $\lambda_2 = \pm \lambda_1 \lambda_3$ and $\lambda_3 = \pm \lambda_1 \lambda_2$.
The BPS equation of the BPS Skyrme model \eqref{LBPS} may also be expressed in terms of these eigenvalues as
\be
\lambda_1\lambda_2 \lambda_3 = \pm \frac{2m}{\lambda} \sqrt{\mathcal{U}(\xi)} .
\ee
This BPS equation implies that the baryon density \eqref{Bden} is always either non-negative or non-positive depending on the choice of sign. This implies that there is no negative baryon density for charge $B>0$ solutions. This can be contrasted with the standard Skyrme model ${\cal L}_{24}$ where negative baryon density has been found in \cite{negBd}. For example for $B=3$ the negative baryon density was found close to the origin and along tubes through the faces of the tetrahedron. Furthermore, it was shown that $\lambda_2^2 \neq \lambda_3^2$ for the $B=3$ Skyrmion.
For the second submodel we have $\lambda_1 = \pm \lambda_2 \lambda_3$ which implies $\lambda_1^2 = \pm \lambda_1 \lambda_2 \lambda_3.$ Hence, in this submodel there is also no negative baryon density for $B>0.$ A similar argument can also be applied to the first submodel, so that all BPS models we discuss here do not have negative baryon density.

%%%%%%%%%%%%%%%%%%%%%%%%%%%%%
\section{ $\mathbb{S}^3$ base space}
%%%%%%%%%%%%%%%%%%%%%%%%%%%%%
Consider the Bogomolny equations (\ref{(1)-bog}) and (\ref{(2)-bog})
on a three dimensional sphere of unit radius with line element
\be
ds^2=d\psi^2+\sin^2\psi d\theta^2 +\sin^2\psi\sin^2\theta d\varphi^2 .
\ee
For the ansatz $\xi = \xi (\psi)$, $u=u(\theta, \varphi)$, then,  equation (\ref{(2)-bog}) becomes
\be
\nabla_\psi \xi = \pm \frac{2i\sin^2\xi}{(1+|u|^2)^2} \left(\nabla_\theta u \nabla_\varphi \bar{u} -
\nabla_\varphi u \nabla_\theta \bar{u}\right),
\ee
where the components of the invariant gradient are given by 
\be
\nabla_\psi = \partial_\psi, \quad 
\nabla_\theta = \frac{1}{\sin \psi} \partial_\theta \quad
{\rm and} \quad 
\nabla_\varphi =  \frac{1}{\sin \psi \sin \theta} \partial_\varphi.
\ee
This is solved by
\be
\xi=\pi - \psi \quad {\rm and} \quad u=\tan \frac{\theta}{2}e^{i\varphi}  \label{sol}
\ee
which has the same boundary conditions as on ${\mathbb R}^3$ if we interpret $\psi = 0$ as the origin and $\psi = \pi$ as ``infinity.'' 
On the other hand, equation (\ref{(1)-bog}) gives
\be
\nabla_\theta u \mp i  \nabla_\psi \xi \nabla_\varphi u=0
\quad {\rm and} \quad
\nabla_\varphi u \pm i  \nabla_\psi \xi \nabla_\theta u=0 .
\ee
This is solved again by (\ref{sol}). Therefore, on $\mathbb{S}^3$ both BPS submodels do lead to a common solution, as expected \cite{sphere}. Solutions of higher topological charge on $\mathbb{S}^3$ are discussed in \cite{sphereB}.

Recently a similar deformation of Skyrme-related Bogomolny equations has been considered, where the coupling constants multiplying the quadratic and quartic terms of the model are replaced by a space dependent function $f$ \cite{Ferreira} (for another possibility see \cite{canf}).  Then the Bogomolny equations take the form
\be 
f^2 u_i\pm i \epsilon_{ijk} \xi_j u_k=0
\ee
and
\be 
f^2 \xi_i \mp  \frac{2i\sin^2 \xi}{(1+|u|)^2} \epsilon_{ijk} u_j \bar{u}_k =0.
\ee
Now, for suitably chosen functions $f$ the resulting Bogomolny equations have common topologically nontrivial solutions on $\mathbb{R}^3$.

Due to the fact that the BPS equations on $\mathbb{S}^3$ base space coincide and have a common solution in the charge one sector we can conclude that solutions of the BPS equations satisfy
\bea
E=E^{(1)}+E^{(2)}&=& \int_{\mathbb{S}^3} d\Omega \left( \lambda^2_2+\lambda_3^2+\lambda_1^2\lambda_3^2+\lambda_1^2\lambda_2^2\right) + \int_{\mathbb{S}^3} d\Omega \left( \lambda^2_1+\lambda_2^2\lambda_3^2\right), \\
&=& 2E^{(2)}+E^{(2)}.
\eea
In other word, the first BPS submodel gives a two times bigger contribution to the total energy than the second BPS submodel for the  $B=1$ soliton solution,
\be
E^{(1)}_{{\rm on-shell}}=2 E^{(2)}_{{\rm on-shell}}
\ee
where the subscript ``on-shell'' emphasises that this is only valid for solutions of the BPS equations. The fate of this relation on $\mathbb{R}^3$ and its relevance for the rational map ansatz will be investigated in Section VII.
%%%%%%%%%%%%%%%%%%%%%%%%%%%%%%%%%%%%%%%%%
\section{$T=0$ thermodynamics of the coupled BPS submodels}
%%%%%%%%%%%%%%%%%%%%%%%%%%%%%%%%%%%%%%%%%
BPS solutions have zero pressure by construction since the energy is topological and, therefore, metric independent \cite{BPS-vort-Brad}. The corresponding BPS equations may be generalised to first-order equations for nonzero pressure, sheding light on the thermodynamical behaviour of the material system described by the solitons.
It is, thus, natural to analyse the soliton solutions in the BPS submodels once a non-zero pressure is imposed.
%%%%%%%%%%%%%%%%%%%%%%%%%%%%%%%%%%%%%%%%%
\subsection{The $\mathcal{L}^{(1)}_{24}$ BPS model and non-zero pressure}
%%%%%%%%%%%%%%%%%%%%%%%%%%%%%%%%%%%%%%%%%
 Static Skyrmions of this model can be found from the ansatz $\xi=\xi(r)$,  together with the rational map ansatz  
\be
u(z)=\frac{p(z)}{q(z)}, \label{RM}
\ee
where $z=\tan \frac{\theta}{2}e^{i\varphi}$ is a stereographic coordinate on the unit sphere $\mathbb{S}^2$ parametrized by the usual angular variables $\theta$ and $\varphi$. 
The resulting reduced energy functional reads
\be
E^{(1)}=4\pi \int dr  \left( 2B \sin^2 \xi (1+\xi'^2) \right) = 4\pi B \int dr  \left( 2 \eta'^2 - 2\eta^2 + 4\eta \right),
\ee
where, for convenience, we have introduced the target space variable
$$
\eta=1-\cos \xi.
$$
Then, the profile function follows from the corresponding reduced Bogomolny equation 
\be
\eta'=\pm \sqrt{\eta(2-\eta)} \label{Bog-L2},
\ee
which has the solution
\be
\eta=1-\cos (\pi -r) \;\; \Rightarrow \;\; \xi=\pi -r
\ee
for $r \in [0,\pi]$ and 0 otherwise. $R=\pi$ is interpreted as the size of the compact Skyrmion. Here we chose the minus sign and imposed the appropriate boundary conditions 
\be
\eta(r=0)=2, \;\; \eta(r=R)=0, \;\; \eta'(r=R)=0. \label{boundary}
\ee 
It is interesting to note that for the BPS submodel $\mathcal{L}_{24}^{(1)}$, all Skyrmions have the same size and volume independently of the value of the topological charge - $R(B)=\pi$ and $V(B)=V_0=\frac{4}{3}\pi^4$. Hence, increasing the baryon charge we increase the energy, $E(B)=8\pi^2|B|$, stored in a fixed volume. Therefore, one can say that this BPS submodel describes a {\it very attractive BPS skyrmionic matter}, where solitons are confined in a fixed volume. The radial energy density is zero both outside $r=R$ and at $r=0$, therefore the individual $B=1$ Skyrmions are distributed on a spherical shell of finite 
thickness which is independent of $B.$ Their angular distribution is given by the rational map $u=(p(z)/q(z))$ of the solution which can be interpreted as the distribution of sigma model lumps on the two-sphere.

The fact that the volume of a soliton and its radial profile function are independent of its topological charge 
somewhat resembles a Bose-Einstein condensate (BEC). In a BEC phase a large fraction of particles occupies the same lowest energy state, described by the same wave function. Adding more particles, which is analogue to increasing the topological charge, just increases the density of the condensate. Furthermore, a BEC is a phenomenon occurring close to $T=0$ which is the relevant phase for the applicability of the Skyrme model.

\vspace*{0.2cm}

The constant volume of Skyrmions together with their BPS nature give a simple expression for the mean-field baryon chemical potential 
\be
\bar{\mu} = \left( \frac{\partial E}{\partial B} \right)_V = 8\pi^2,
\ee
which is equal to the energy of the charge one Skyrmion. 

Furthermore, due to the compacton nature of Skyrmions in this submodel, there is another phase of Skyrmionic matter - a gas of $N$ non-overlapping  $B=1$ Skyrmions, each of volume $V=V_0$. This phase has exactly the same energy as the charge $B=N$ Skyrmion, but the total volume of the configuration is now $N$ times bigger. This should be contrasted with liquid and gas phases 
in the BPS Skyrme model where the volume and energy of a soliton are always linear functions of the baryon charge. 

\vspace*{0.2cm}

The second order Euler-Lagrange equation for the profile $\xi$
\be
\eta''+ \eta - 1=0
\ee
is solved not only for the Bogomolny equation (\ref{Bog-L2}) but also for a whole family of first order equations parametrised by a parameter $C,$ namely
\be \label{Bog-P-L2}
\eta'^2= \eta(2-\eta) +\frac{C}{2B} .
\ee
This equation can be analytically solved providing the squeezed Skyrmion solutions
\be
\eta(r) = \left\{
\begin{array}{ll}
1- \frac{\sin  \left(r-\frac{R}{2} \right)}{\sin \frac{R}{2}}  & \;\;\;\; r \leq R, \\
& \\
0 &\;\;\;\; r \geq R,
\end{array}
\right.
\ee
where the size of the Skyrmion is 
\be
R=2 \arctan \sqrt{\frac{2B}{C}},
\ee
and the volume $V$ satisfies the useful identity
\be
\label{usefulV}
\tan^2 \frac{1}{2}  \left( \frac{3V}{4\pi} \right)^{1/3} = \frac{2B}{C}.
\ee
The parameter $C$ measures the squeezing rate of the solution and therefore is related to the external pressure imposed on the original BPS solution at zero pressure. The energy of this solution for general $C$ is
\be
E^{(1)}(P)= 4\pi B  \int_0^R dr  \left( 2 \eta'^2 - 2\eta^2 + 4\eta \right) 
= 16 \pi B  \int_0^2 d\eta \frac{2\eta-\eta^2 +\frac{C}{4B}}{\sqrt{2\eta-\eta^2 +\frac{C}{4B}}},
\ee
where the first order equation (\ref{Bog-P-L2}) has been used to transform the base space integral into a target space integral. Then we find
\be
E^{(1)}(C)=16\pi B \left( \sqrt{\frac{C}{2B}} +\arctan \sqrt{\frac{2B}{C}}\right),
\ee
which can be written with \eqref{usefulV} in terms of the volume as
\be
E^{(1)}(V)=16\pi B \left( \frac{1}{ \tan  \left( \frac{3V}{32\pi}  \right)^{1/3} } + \left( \frac{3V}{32\pi}  \right)^{1/3} \right) .
\ee
This expression is linear in $B$, therefore, the mean-field baryon chemical potential is again equal to the energy of the $B=1$ Skyrmion, now at nonzero pressure. Furthermore, this energy allows us to compute the proper thermodynamical pressure
\be
p=-\frac{\partial E(V)}{\partial V} = \frac{C}{\left( 2\arctan \sqrt{\frac{2B}{C}} \right)^2}.
\ee
In other words the parameter $C$ gives, in a rather complicated way, the thermodynamical pressure, namely
\be
p=\frac{C}{R^2(C)},
\ee
where we explicitly use the formula for the size of the squeezed Skyrmion $R(C)$. In the small volume limit (large $C$ parameter) the energy and the pressure take the form
\be
E^{(1)}(V)=16\pi B  \left( \frac{32\pi}{3V}  \right)^{1/3}
\quad {\rm and}
\quad p=16\pi B  \left( \frac{32\pi}{3V}  \right)^{1/3} \frac{1}{3V}.
\ee
This leads to the expected high pressure limit of the mean-field equation of state relating the pressure and the mean-field energy density $\bar{\rho} = E/V,$ namely
\be
p=\frac{\bar{\rho}}{3}.
\ee
This is exactly the mean-field 
%EoS 
equation of state
of the $\mathcal{L}_{24}$ Skyrme model \cite{EoS}, \cite{Kut}.
%%%%%%%%%%%%%%%%%%%%%%%%%%%%%%%%%%%%%%%%%
\subsection{The $\mathcal{L}^{(2)}_{24}$ BPS model and non-zero pressure}
%%%%%%%%%%%%%%%%%%%%%%%%%%%%%%%%%%%%%%%%%

Within this submodel it is not possible to simultaneously impose both boundary conditions for the profile function, $\xi(r=0)=\pi$ and $\xi(r=\infty)=0,$ because the condition $\xi (r=\infty) =0$ is not required for finite energy. Therefore, the Skyrmion solutions do not possess integer baryon charge. 
An interpretation is that there are too strong repulsive forces in this submodel, such that a Skyrmion cannot form, as opposed to the compactons in the $\mathcal{L}^{(1)}_{24}$ model. 
Acting with an additional external force by applying external pressure should give rise to conventional Skyrmions.
Let us start with the static energy for the second BPS submodel, where we insert $\xi = \xi (r)$ and the solution for the complex field $u=\tan \frac{\theta}{2} e^{iB\varphi},$  resulting in
\be
E^{(2)}= 4\pi \int_0^\infty dr r^2 \left(\xi'^2+\frac{B^2\sin^4 \xi}{r^4}  \right). \label{E2}
\ee
It is convenient to introduce the new base space variable $y=1/r$, giving
\be
E^{(2)}= 4\pi \int_0^\infty dy  \left(\xi_y^2+B^2\sin^4 \xi  \right).
\ee
Again, the full second order Euler-Lagrange equation for $\xi$ is solved not only by the Bogomolny equation but also by its one-parameter $(D\geq0)$ generalisation
 \be
 \xi_y^2=B^2\sin^4 \xi +D, \label{P-E2}
 \ee
where $D=0$ gives the Bogomolny equation. The non-zero pressure boundary conditions translate as 
 \be
 \xi(y=\infty)=\pi, \quad \xi(y=y_0)=0, \quad {\rm and} \quad \xi_y(y=y_0)=0,
 \ee
where $y_0=R^{-1}$ and $R$ is a compacton boundary at which we impose an external pressure. But the first condition leads to difficulties. Namely, at leading order at $y\rightarrow \infty$, $\xi_y^2=D$. As a consequence, the formal solution $\xi=\pm \sqrt{D}y$ is unbounded which contradicts the assumed condition at $y\rightarrow \infty,$ namely $\xi = \pi$. Therefore, 
the $y=\infty$ $(r=0)$ boundary condition 
cannot be satisfied for solutions of the non-zero pressure $(D>0)$ equation (\ref{P-E2}). 
 
It is instructive to recall the BPS case with no squeezing and $D=0,$ where
the boundary condition $\xi(r=0)=\pi$ can be imposed but $\xi(r=\infty) \neq 0.$
Hence, qualitatively the squeezing brings the $\xi = 0$ end from ``beyond infinity'' to a finite distance while the solution at the origin diverges.

The problem with $D>0$ becomes more transparent if we insert the generalization of the Bogomolny equation to the total energy so that
\be
E^{(2)}= 4\pi \int_0^\infty dy  \left( 2 B^2\sin^4 \xi +D  \right).
\ee
Obviously, for $D>0$ the second term leads to infinite energy at $y=\infty$ which is the origin $r=0.$ Hence, in order to squeeze such a skyrmionic matter 
we have to use an infinite amount of energy or act with infinite pressure. 
We interpret this as a {\it very repulsive BPS skyrmionic matter} which cannot be squeezed by finite pressure. 

\vspace*{0.2cm}

A Skyrmion cannot exist in a ball of finite volume because of the singular behaviour at $r=0$. This can be resolved by also ``squeezing'' the configuration from the inner region which is achieved by the following boundary conditions
  \be
 \xi(r=R_1)=\pi \quad {\rm and} \quad  \xi(r=R_2)=0
 \ee
or equivalently, 
 \be
\xi(y=R_2^{-1})=0 \quad {\rm and} \quad \xi(y=R_1^{-1})=\pi.
 \ee
One can easily verify that equation (\ref{P-E2}) has finite energy solutions satisfying such boundary conditions. These solutions may be expressed in terms of hypergeometric functions, but the resulting expressions are rather complicated and not very instructive, so we do not show them here. An instructive example can be provided in the limit when $D \gg B^2$ which, physically, corresponds to the limit of high pressure and high density. Then, we can choose the plus sign in (\ref{P-E2}) and obtain 
\be
\xi_y=\sqrt{D},
\ee
leading to the  solution with baryon charge $B$ 
\be
\xi(r)=
\sqrt{D} \left(  \frac{1}{r} - \frac{1}{R_2} \right),
\ee
where $r\in [R_1,R_2]$. Furthermore the radii are related by the following condition 
\be
\pi = \sqrt{D} \left(  \frac{1}{R_1} - \frac{1}{R_2} \right) .
\ee
The corresponding energy reads
\be
E=4\pi D  \left(  \frac{1}{R_1} - \frac{1}{R_2} \right) = 4\pi^3 \frac{R_1R_2}{R_2-R_1},
\ee
and the volume of the solution is
\be
V = \frac{4\pi}{3} (R_2^3 - R_1^3).
\ee
It follows that the energy cannot be expressed solely by the volume, but depends separately on the volume and on the ``size'' (e.g. $R_2$) of the solution which is related to the fact that the underlying field theory is not of the perfect fluid type. For such field theories, the correct thermodynamical definition of the pressure is given by the Weyl rescaling of the energy functional, see \cite{BPS-vort-Brad}. If the space coordinates
in $d$-dimensional Euclidean space are rescaled by  $\vec x \to e^{\lambda}\vec x$, then the pressure is given by
\be
p = \left. \frac{1}{dV}\frac{\partial E}{\partial \lambda}\right|_{\lambda =0}.
\ee
For the above energy expression the Weyl rescaling is just $R_i \to e^\lambda R_i$, leading to the pressure and equation of state
\be
p = \frac{1}{3V} E \; , \quad \bar \rho \equiv \frac{E}{V} \; \quad \Rightarrow \quad p = \frac{\bar \rho}{3},
\ee
which is the expected equation of state in the limit of high density. 
%%%%%%%%%%%%%%%%%%%%%%%%%%%%%%%%%%%%%%%%%
\subsection{Oscillons in the $\mathcal{L}^{(1)}_{24}$ BPS model}
%%%%%%%%%%%%%%%%%%%%%%%%%%%%%%%%%%%%%%%%%
Although this issue is somewhat outside the main line of the present paper, it is interesting to observe that in the $\mathcal{L}^{(1)}$ BPS submodel there exists a different type of non-topological and non-static soliton, the so-called oscillon. We first observe that the ansatz $\xi=\xi(r,t)$, $u=u(\theta, \varphi)$ is still compatible with the field equations and $u$ continues to be solved by rational maps $u(z)$ for this ansatz. 
\\
In order to prove it we note that the ansatz implies the orthogonality $\xi_\mu u^\mu=\xi_\mu \bar{u}^\mu\equiv0$. As a consequence,
 the last term in the Lagrangian density (\ref{L(1)}) vanishes identically and, as it is quadratic in the action, it also vanishes in the equations of motion. Hence, for the above ansatz, the model is just the $CP^1$ model multiplied by a real scalar field model, with Lagrangian density 
 \be
 \mathcal{L}_{24}^{(1)} = \mathcal{L}_{CP^1} \mathcal{L}_\xi \; , \quad 
 \mathcal{L}_{CP^1} = \frac{4u_\nu \bar{u}^\nu}{(1+|u|^2)^2} \; , \quad 
 \mathcal{L}_\xi = \sin^2\xi  (1-  \xi_\mu \xi^\mu ) .
 \ee
 The ansatz implies for the Euler-Lagrange (EL) variation of $u$
 that
 \be
 \left( \frac{\partial }{\partial u} - \partial_\mu \frac{\partial}{\partial u_\mu} \right) \mathcal{L}_{24}^{(1)}
 = \mathcal{L}_{\xi} \left( \frac{\partial }{\partial u} - \partial_\mu \frac{\partial}{\partial u_\mu} \right)
 \mathcal{L}_{CP^1}
 \ee
 because $\mathcal{L}_\xi$ only depends on $r$ and $t$. Hence, the EL equations are just the field equations for the $CP^1$ model. For the variation w.r.t. $\xi$ we use that
 $\mathcal{L}_{CP^1} = r^{-2} \tilde{\mathcal{L}}_{CP^1} (\theta ,\varphi)$ and find
 \be
 \left( \frac{\partial }{\partial \xi} - \partial_\mu \frac{\partial}{\partial \xi_\mu} \right) \mathcal{L}_{24}^{(1)}
 = r^{-2}\tilde{\mathcal{L}}_{CP^1} \left( \frac{\partial }{\partial \xi} - \partial_\mu \frac{\partial}{\partial \xi_\mu} + \frac{2}{r} \frac{\partial}{\partial \xi_r} \right)
 \mathcal{L}_\xi
 \ee
 where the only effect of the last term is to replace the three-dimensional radial Laplacian $\partial^2_r + (2/r) \partial_r$ by the one-dimensional Laplacian $\partial^2_r$. 
To find the equivalent symmetry-reduced model for the ansatz we now separate the Lagrangian
$L_{24}^{(1)} \;\; = \; \; \int d\Omega_{\mathbb{R}^3} \mathcal{L}_{24}^{(1)}$ as 
\be
L_{24}^{(1)} \;\; = \; \; \left( - 2\int \Omega_{\mathbb{S}^2}  \frac{(1+z\bar{z})^2}{(1+u\bar{u})^2} (u_z\bar{u}_{\bar{z}} +u_{\bar{z}}\bar{u}_z) \right) \int dr \sin^2\xi \left( 1  -  \xi_\mu \xi^\mu \right)
\ee
where we used that
\be
d\Omega_{\mathbb{R}^3}=dr r^2 d\Omega_{\mathbb{S}^2}, \;\;\;  d\Omega_{\mathbb{S}^2}=\frac{2i}{(1+|z|^2)^2} dzd\bar{z}
\ee
and 
\be
u_\mu \bar u^\mu=-u_i \bar{u}_i = -\frac{(1+z\bar{z})^2}{2r^2} (u_z\bar{u}_{\bar{z}} +u_{\bar{z}}\bar{u}_z) .
\ee
Here we introduced the stereographic coordinate $z=\tan \frac{\theta}{2} e^{i\varphi}$. Note that the $r^{-2}$ factor from $u_\mu u^\mu $ cancels with the $r^2$ factor from the volume form. The $CP^1$ part is minimised by rational maps of degree $B$ with energy $E_{CP^1}=4\pi |B|$. Hence, 
\be
L_{24}^{(1)} \;\; = \; \; -2E_{CP^1} \int dr \sin^2\xi \left( 1  -  \xi_\mu \xi^\mu \right) = 8\pi |B| \int dr \sin^2\xi \left(  \xi_\mu \xi^\mu -1 \right)   
\ee
or, using $\eta = 1-\cos \xi$,
\be
{L}^{(1)} = 8\pi B \int dr \left( \eta_\mu \eta^\mu- \eta(2-\eta)\right) \label{model}
\ee
where the r.h.s. is formally equivalent to a scalar field theory in 1+1 dimensions with a potential with two vacua.
To investigate solutions with small amplitudes around the first vacuum $\eta =0$ we substitute $\eta (t,r)=\epsilon (t,r) \geq 0$ and find
\be
{L}^{(1)}=8\pi B \int dr \left(\epsilon_\mu \epsilon^\mu- 2\epsilon \right).
\ee
The resulting equation of motion is 
\be
\partial_t^2 \epsilon (t,r) - \partial_r^2 \epsilon (t,r) =-1 .
\ee
Since the perturbation cannot take negative values one has to specify what happens for $\epsilon\rightarrow 0$. 
Following \cite{arodz1} we equip the field equation with the elastic bounce condition at $\epsilon(t,r)=0$  relating the field velocities before and after bouncing. Namely
\be
\partial_t \epsilon (t,r) \rightarrow - \partial_t \epsilon (t,r) \;\;\;\;  \mbox{when} \;\;\;\; \epsilon=0.
\ee 
This condition can be removed if we extend the field (the target space) to a new auxiliary field (extended target space) $\tilde{\epsilon} \in \mathbb{R}$ 
with $\epsilon(t,r) = |\tilde{\epsilon}(t,r)|$
by performing the unfolding procedure as described in \cite{arodz1}. As a consequence, we derive the following evolution equation
\be
\partial_t^2 \tilde{\epsilon} (t,r) - \partial_r^2 \tilde{\epsilon} (t,r) =- \mbox{sign} \left( \tilde{\epsilon} (t,r) \right),
\ee
 which is 
the signum-Gordon equation in (1+1) dimension. Strictly speaking, it is a version of the model on $\mathbb{R}\times \mathbb{R}_+$. An interesting observation is that this equation still admits breather-like solutions which are stable, non-radiating and time-periodic \cite{arodz2}. Furthermore, these solutions are known in an exact form \cite{arodz2}. Let
 \be
\tilde{\epsilon}_1(t,r) = \left\{
 \begin{array}{lll}
 -\frac{r^2}{2} & &0\leq r \leq t, \\
 \frac{t^2}{2}-rt & &t\leq r \leq \frac{1}{2}-t, \\
 \frac{r^2}{2}+t^2 -\frac{r}{2}-\frac{t}{2}+\frac{1}{8} &  &\frac{1}{2}-t\leq r \leq \frac{1}{2}+t, \\
 \frac{t^2}{2}+t(r-1) &  &\frac{1}{2}+t\leq r \leq 1-t, \\
 -\frac{(1-r)^2}{2} & & 1-t\leq r\leq 1,\\
0 & & {\rm otherwise,}
 \end{array}
 \right.
 \ee
for $t \in \left[0,\frac{1}{4}\right]$ and 
\be
\tilde{ \epsilon}_2(t,r) = \left\{
 \begin{array}{lll}
 -\frac{r^2}{2} & & 0\leq r \leq \frac{1}{2}-t, \\
 \frac{t^2}{2}+tr-\frac{r}{2}-\frac{t}{2}+\frac{1}{8} & & \frac{1}{2}-t\leq r \leq t, \\
 \frac{r^2}{2}+t^2 -\frac{r}{2}-\frac{t}{2}+\frac{1}{8} & & 1- t\leq r \leq \frac{1}{2}+t, \\
 \frac{t^2}{2}-tr +\frac{r}{2}+\frac{t}{2}-\frac{3}{8} & &1-t\leq r \leq \frac{1}{2}+t, \\
 -\frac{(1-r)^2}{2} & &\frac{1}{2}+t\leq r\leq 1,\\
0 & & {\rm otherwise,}
 \end{array}
 \right.
 \ee
for $t\in \left[\frac{1}{4},\frac{1}{2}\right]$. 
Then the solution for time $t$ can be written as
\be
\tilde{\epsilon}(t,r) = \left\{
\begin{array}{lll}
\tilde{\epsilon}_1(t,r) & & 0 \le t \le \tfrac{1}{4},\\
\tilde{\epsilon}_2(t,r) & & \tfrac{1}{4} \le t \le  \tfrac{1}{2}, \\
-\tilde{\epsilon}_1(t-\tfrac{1}{2}) & & \tfrac{1}{2} \le t \le \tfrac{3}{4},\\
-\tilde{\epsilon}_2(t-\tfrac{1}{2}) & & \tfrac{3}{4} \le t \le 1,\\
\tilde{\epsilon}(t,r) = \tilde{\epsilon}(t+1,r) & & {\rm otherwise.}
\end{array}
\right.
\ee
This solution has period $T=1$ and describes an oscillating shell of size $R=1$ and with the center at $R_c=\frac{1}{2}$. Using the translation invariance of the reduced model it can be trivially moved to any position $R_c>\frac{1}{2}$.  

Moreover, since the signum-Gordon equation is dilatation invariant, the breather solution constitutes in fact an infinite family of solutions
\be
\tilde{\epsilon}_l(t,r)=l^2\tilde{\epsilon} \left(\frac{t}{l}, \frac{r}{l} \right) ,
\ee
where the arbitrary parameter $l$ is the size and the period of the solution. The amplitude is $\frac{l^2}{16}$ and the energy is
\be
E=\frac{2}{3} \pi B l^3 .
\ee 
Of course, our assumption of small amplitude leads to a restriction on the parameter $l\ll1$. 

To summarise, these solutions are approximate solutions, 
and therefore the true solutions are not breathers but very long lived oscillons. Analogous long lived oscillons can also be found for small field fluctuations about the second vacuum at $\eta=2$. 

Compact breathers with arbitrarily small amplitude (arbitrarily long lived compact oscillons in the model (\ref{model})) have arbitrarily small energy and therefore form a sort of an infrared cloud (a composition of non-overlapping compactons) which may dominate radiation/interaction in the model (\ref{model}). 
It is also worth emphasizing that the oscillons are genuine 3+1 dimensional non-topological objects (non-topological shell solitons) even though described by the effective 1+1 dimensional signum-Gordon equation.   
The detailed analysis of the oscillons in the model (\ref{model}), their fate in the $\mathcal{L}^{(1)}_{24}$ BPS model as well as their role in the full Skyrme model is very interesting but goes beyond the scope of the present work. 
The perturbation of the signum-Gordon model by a quadratic part of the potential has been investigated in \cite{klimas}. 
%Let us remark that 
In addition to breathers with a fixed boundary, the signum-Gordon model also gives rise to breathers with oscillating boundaries \cite{swier}.  Consequently, the model (\ref{model}) should contain oscillons with oscillating boundaries (inner and outer radial boundary) which become long lived in the limit of small amplitude. 

Finally, nontopological long-lived breather-like structures in the Skyrme model have been reported \cite{zak osc}. It would be 
interesting to verify if they are related with the presented signum-Gordon breathers of the $\mathcal{L}^{(1)}_{24}$ submodel. 
%%%%%%%%%%%%%%%%%%%%%%%%%%%%%%%%%%%%%%%%%
\section{The $\mathcal{L}^{(2)}_{24}$ BPS model and its solvable non-BPS extension}
%%%%%%%%%%%%%%%%%%%%%%%%%%%%%%%%%%%%%%%%%
As we know, the second coupled BPS submodel does not support Skyrmions with an integer baryon number.
In fact, this may be interpreted as a strong repulsion built into the model. 
%%Another interpretation is that Skyrmions have ``more than infinite'' size, and they do not fit into $\mathbb{R}^3$. 

The addition of a potential
\be
\mathcal{L}=\mathcal{L}^{(2)}_{24}+m^2 \mathcal{L}_0
\ee
 not only breaks the BPS property of this submodel 
but also increases the attractive force acting on the Skyrmion. This {\it may} result in the appearance of the usual 
infinitely extended Skyrmions which possess an integer valued baryon charge. 

The first important observation is that the ansatz assumed for  $\mathcal{L}^{(2)}_{24}$ BPS submodel 
\be
\xi=\xi(r) \quad {\rm and} \quad u=v(\theta) e^{iB\varphi},
\ee
still works and gives $v=\tan \frac{\theta}{2}$ and the radial energy functional
\be
E^{(2)}+E_0= 4\pi \int dr r^2 \left(\xi'^2+\frac{B^2\sin^4 \xi}{r^4}  + m^2 \mathcal{U}(\xi)\right) \label{L2+L0},
\ee
where $E_0= 4\pi m^2 \int dr r^2  \mathcal{U}(\xi)$ is the contribution from the potential part. This ansatz continues to work even after 
including the usual BPS Skyrme term. 
A finite energy requirement is that  $\xi (r=0)= n\pi$, for
$n\in \mathbb{Z}$ (we chose $\xi (r=0)=\pi$). 

In the following we assume $\mathcal{U}(\xi=0)=0,$ so that $\mathcal{U}$ has its vacuum at $\xi=0.$ 
Then the second boundary condition is $\lim_{r\to \infty} \xi (r)=0$. 
Using these two boundary conditions,
the lower topological energy bound becomes 
\be
E^{(2)}+E_0\geq E^{(2)} \geq 4\pi^2 |B|. \label{E20_bound}
\ee
%%%%%%%%%%%%%%%%%%%%%%%%%%%%%%%%%%%%%%%%%
\subsection{The pion mass potential $\mathcal{U}_\pi$}
%%%%%%%%%%%%%%%%%%%%%%%%%%%%%%%%%%%%%%%%%
To find solutions, when a pion mass term is included, we need to consider the energy functional
 \be
E^{(2)}+E_0= 4\pi \int dr r^2 \left(\xi'^2+\frac{B^2\sin^4 \xi}{r^4}  + m^2(1-\cos \xi) \right). %\label{L2-pion}
\ee
One can redefine the radial coordinate $r\rightarrow B r$ to obtain a one parameter family of models with the energy scale 
multiplied by the charge $B$
\be
E^{(2)}+E_0= 4\pi B  \int dr r^2 \left(\xi'^2+\frac{\sin^4 \xi}{r^4}  + \beta^2 (1-\cos \xi) \right) \label{L2-pion}
\ee
and $\beta^2 \equiv B^2m^2$. The corresponding field equation is
\be \label{field-eq}
-2\partial_r (r^2\xi') +\frac{4}{r^2} \sin^3\xi \cos \xi +\beta^2 r^2\sin \xi=0.
\ee
Expansion at the origin where $\xi =\pi-\eta +o(\eta)$ is governed by the first two terms and is not affected by the potential. We find that
\be
\xi=\pi -r +o(r).
\ee
On the other hand, at $r\rightarrow \infty$  where $\xi=\eta +o(\eta)$ we obtain
\be
\xi=Ae^{-\frac{\beta}{\sqrt{2}}r}.
\ee
The existence of the expansions at $r = 0$ ($\xi=\pi$) and at $r=\infty$ ($\xi=0$) gives some evidence that there can exist integer baryon charge Skyrmions for (\ref{L2-pion}). Especially, if one compares with what happens for the BPS case without potential  when
\be
-2\partial_r (r^2\xi') +\frac{4}{r^2} \sin^3\xi \cos \xi =0.
\ee
For the asymptotic expansion at infinity we assume $\eta=Cr^\alpha,$ which leads to
\be
-2A\alpha(\alpha+1)r^\alpha+4A^3r^{3\alpha-2}=0.
\ee
Hence, $\alpha=1$. But this contradicts our assumption that $\xi$ (or $\eta$) is close to the vacuum value for $r\rightarrow \infty$. So, there is no expansion at infinity which would give $\xi=0$ for $r=\infty$. This completely agrees with our previous finding that there is no integer baryon charge Skyrmions for the $\mathcal{L}^{(2)}$ BPS submodel.

\begin{figure}
\hspace*{-1.0cm}
\includegraphics[height=6.cm]{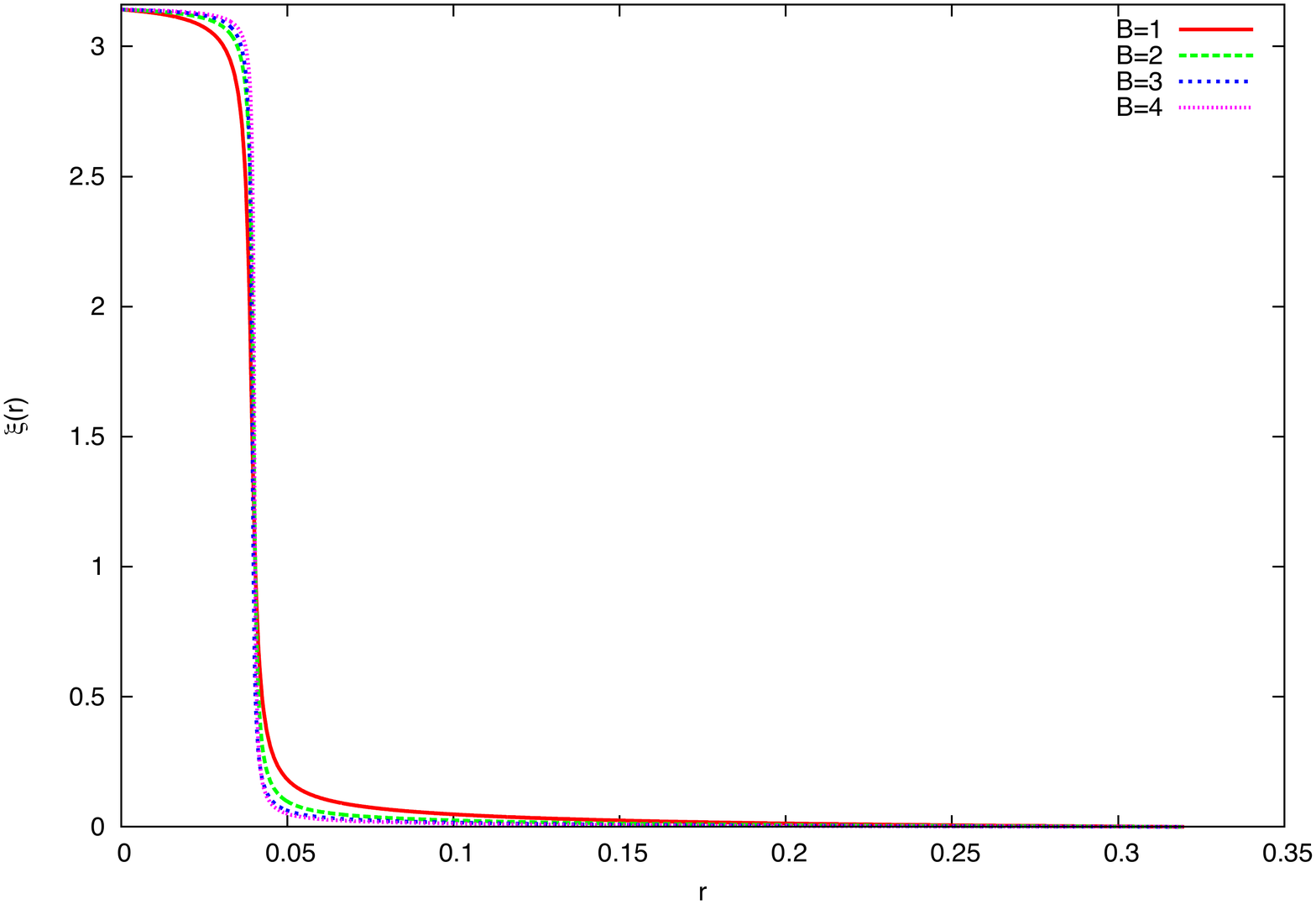}
\includegraphics[height=6.cm]{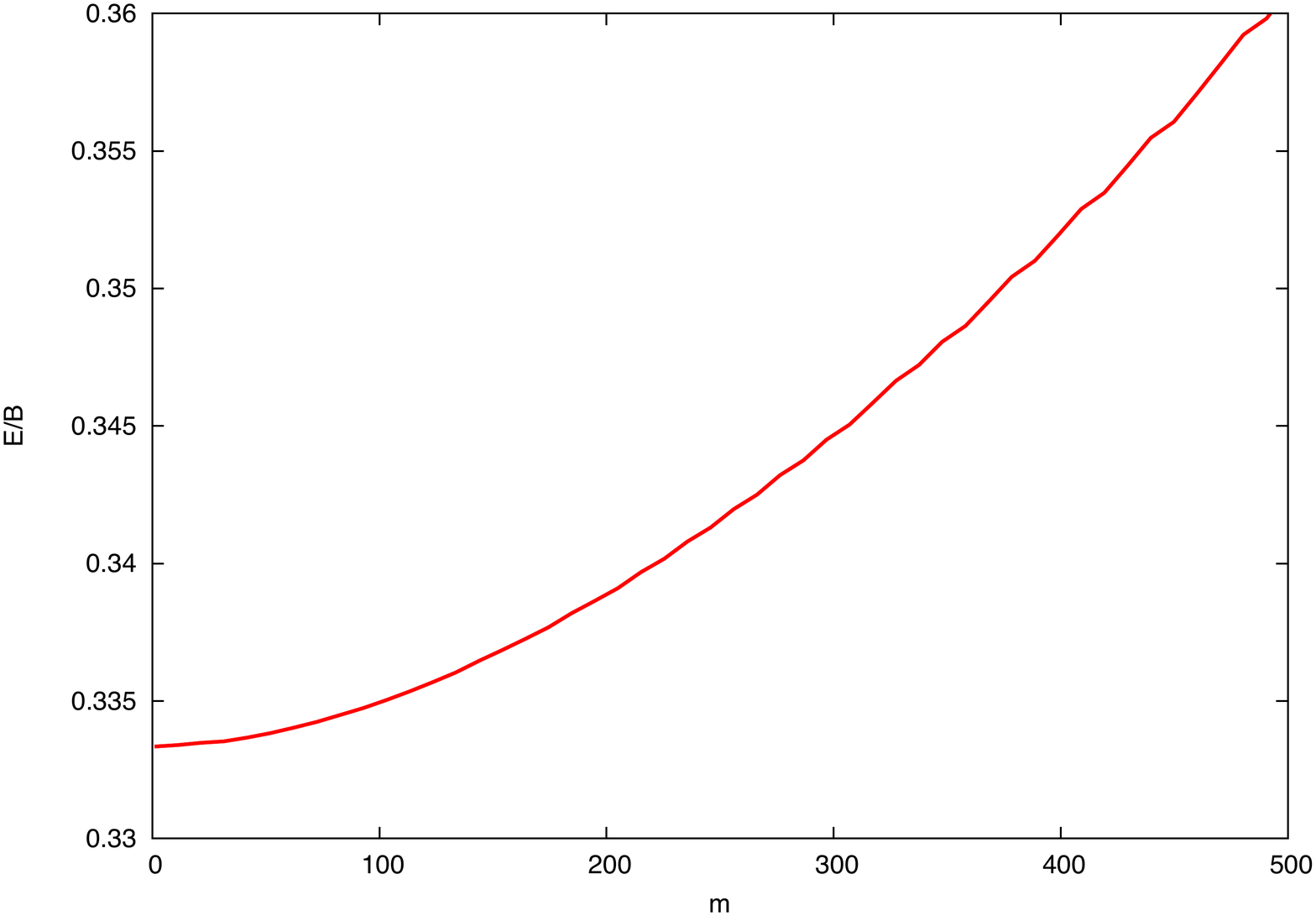}
\caption{Left: Profile function $\xi$ in the $E^{(2)}+E_0$ model with the pion mass potential and $m=1$. Right: Energy divided by $12 \pi^2 |B|$  as a function of the mass parameter for $B=1$ (there is no discernible difference for $B=2,3,4$).} 
\label{L2-fg}
\end{figure}

Therefore, to solve the differential equation \eqref{field-eq} we need to proceed numerically. We approximate the derivatives by fourth order finite differences on a numerical lattice, and minimise the 
energy functional with gradient flow. This produces an artificial solution which is supported by the numerical lattice and the solution shrinks as the lattice spacing is reduced. This is due to what numerically seems to be an infinite derivative. To proceed we consider the inverse problem. This is analogous to solving a differential equation by separation of variables. We make use of the identity $\xi^{-1}\left(\xi(r)\right)=r$ to rewrite the differential equation \eqref{field-eq} as ($\dot r \equiv dr/d\xi$),
\begin{align}
-4r\dot{r}^2+2r^2\ddot{r}+\frac{4B}{r^2}\dot{r}^3\sin^3\xi\cos\xi+\dot{r}^3\beta^2r^2\sin\xi=0.
\end{align} 
We now consider the radius as $r(\xi)$, with the boundary conditions $r(\xi=0)=\infty$ and $r(\xi=\pi)=0$. Solving this, with gradient flow, produces the image in Fig. \ref{L2-fg}.
As a further check we placed the solution into the gradient flow for $\xi(r)$ with a lattice spacing of $0.000001$ and verified the previous solutions. 

The Skyrmion solutions we obtain have rather remarkable qualitative features. First of all, the profile $\xi$ looks like a step-function. Secondly the size of the Skyrmion, here identified with the position of the rapid jump of $\xi,$ does not significantly vary 
as we change the baryon charge. The corresponding energy is very close to the bound (\ref{E20_bound}). For $m=1$ we find $E/(12\pi^2 B)=0.3333$ for $B=1,2,3,4$, which agrees with the bound for the given numerical precision. Furthermore, the energy grows only very slowly as we increase the mass parameter and go away from the BPS regime as shown in Fig. \ref{L2-fg}.  For $m\approx 500$ we found  $E/(12\pi^2 B)=0.36$. 

\begin{figure}
\hspace*{-1.0cm}
\includegraphics[height=7.cm]{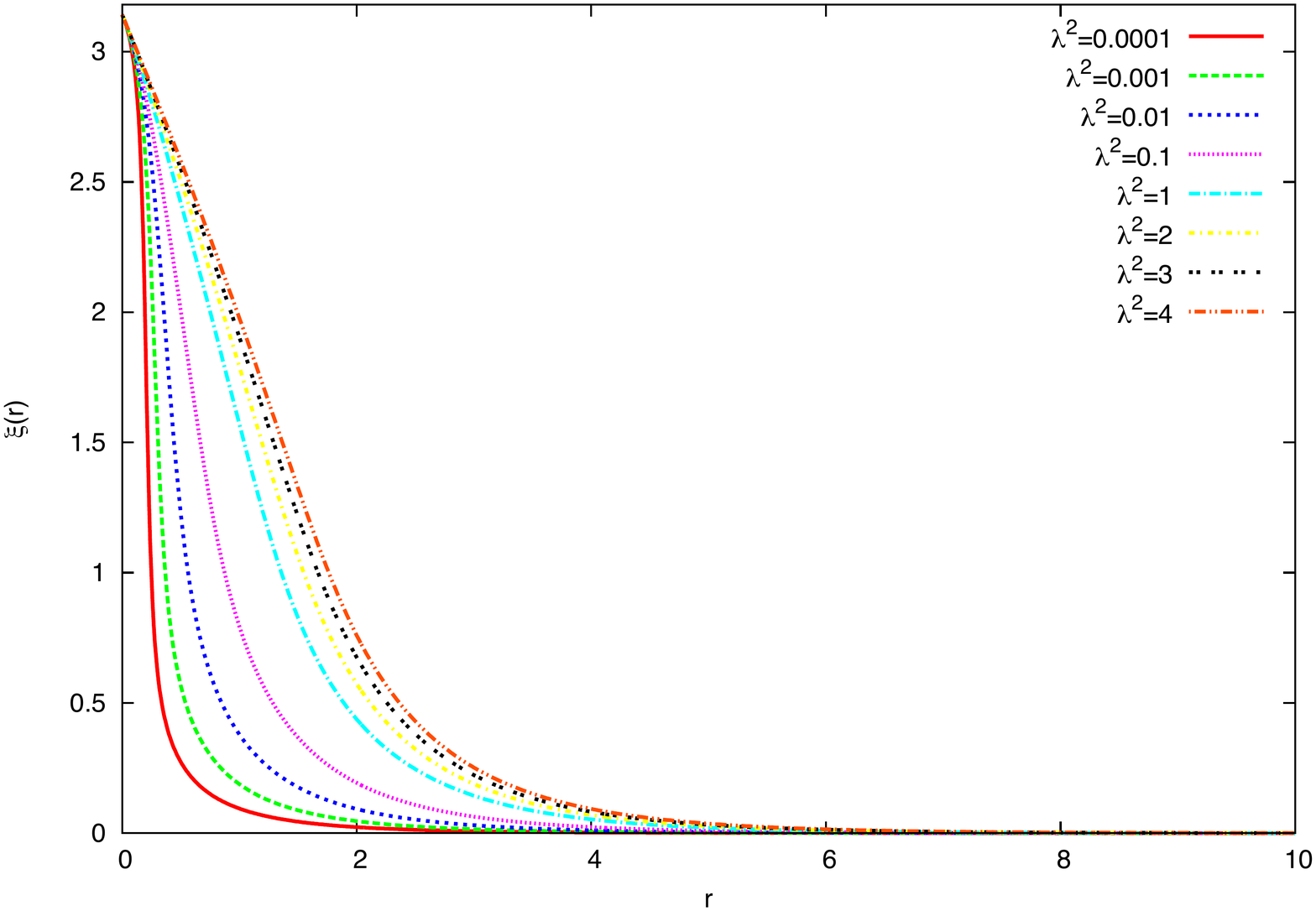}
\caption{Profile function $\xi$ in the $\mathcal{L}^{(2)}+\mathcal{L}_{BPS}$ in the $B=1$ sector. Here we assume the pion mass potential $\mathcal{U}_\pi$ with $m=1$.} 
\label{L2L_BPS-xi}
\end{figure}
\begin{figure}
\hspace*{-1.0cm}
\includegraphics[height=6.cm]{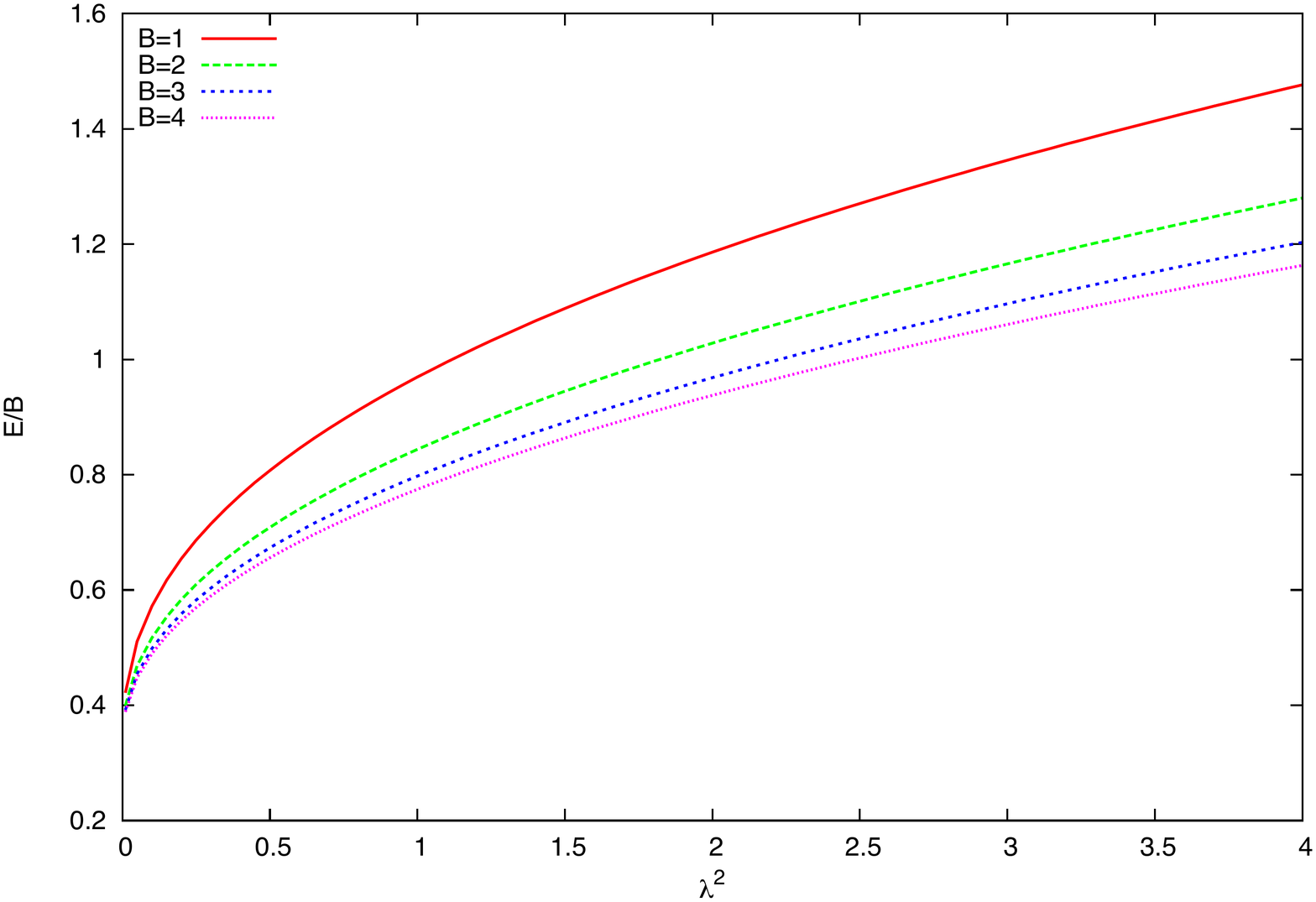}
\includegraphics[height=6.cm]{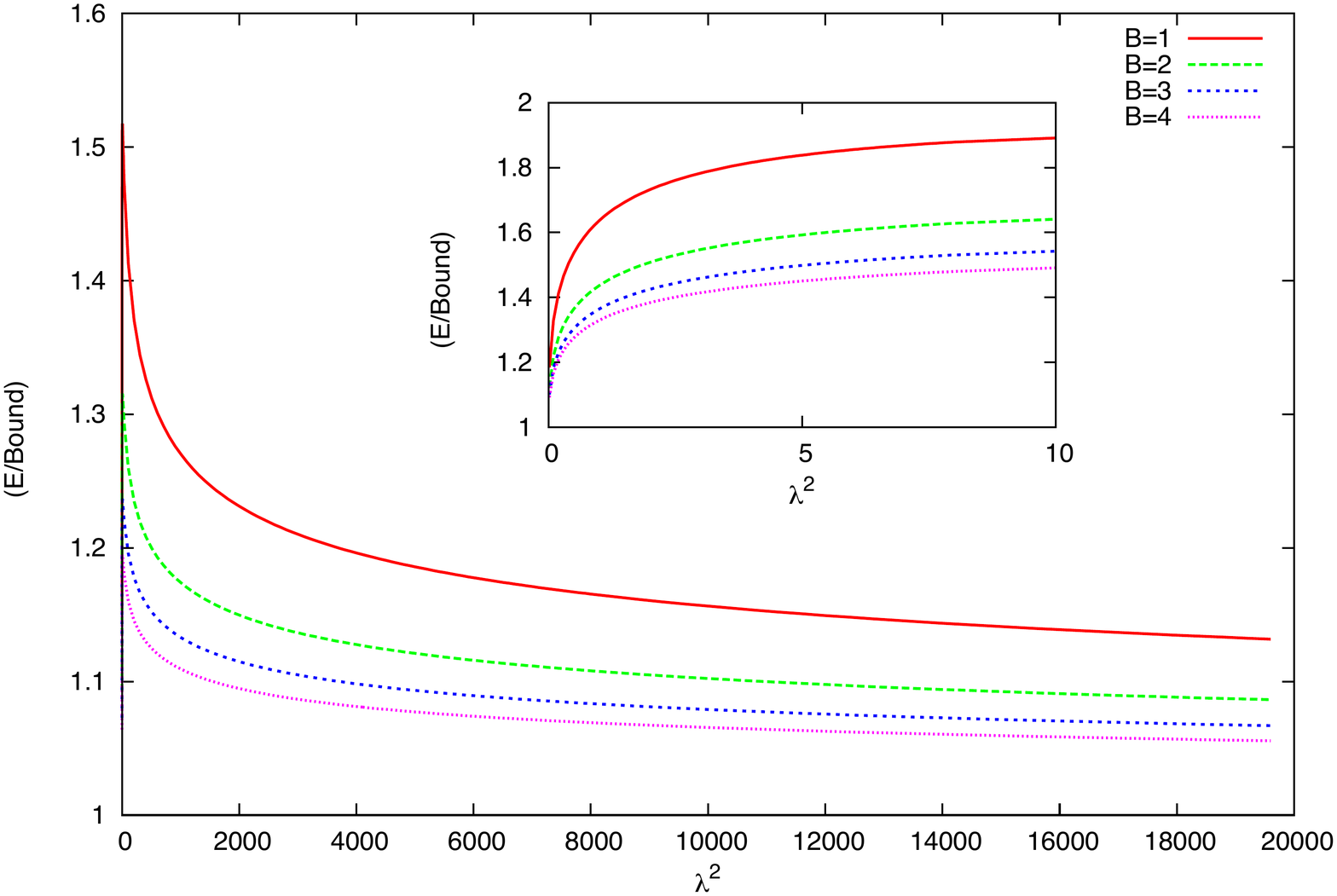}
\caption{Energy per baryon charge (left) and energy divided by the topological bound (\ref{E2+BPS-bound}) (right) for $B=1,2,3,4$ in the $\mathcal{L}^{(2)}+\mathcal{L}_{BPS}$ model, as a function of the coupling constant $\lambda$. Here we assume the pion mass potential $\mathcal{U}_\pi$ with $m=1$.} 
\label{L2L0+L6}
\end{figure}
%%%%%%%%%%%%%%%%%%%%%%%%%%%%%%%%%%%%%%%%%
\subsection{Inclusion of the sextic term}
%%%%%%%%%%%%%%%%%%%%%%%%%%%%%%%%%%%%%%%%%
As we have already mentioned, adding the sextic term does not spoil the applicability of the 
ansatz. This is important, since it allows us to reduce the problem 
%of finding the Skyrmion 
to a second order ODE for the profile function $\xi$. In fact, $\mathcal{L}_0$ and $\mathcal{L}_6$ constitute the BPS Skyrme model, which for the assumed ansatz reads
\be
E_{BPS}=4\pi \int dr  r^2 \left(  \frac{\lambda^2 B^2 \sin^4 \xi \xi'^2}{4r^4}  + m^2 \mathcal{U}(\xi) \right).
\ee
Here we want to analyze the existence and properties of Skyrmions in a model which is a sum of the second BPS submodel and the BPS Skyrme model
\be
E^{(2)}+E_{BPS}= 4\pi \int dr r^2 \left(\xi'^2+\frac{B^2\sin^4 \xi}{r^4} +\frac{\lambda^2 B^2 \sin^4 \xi \xi'^2}{4r^4}  +m^2 \mathcal{U} \right), \label{L2+L_BPS}
\ee
where, for simplicity, the potential is chosen as the standard pion mass potential $ \mathcal{U}=m^2(1-\cos \xi)$. Equivalently, one can treat the model (\ref{L2+L_BPS}) as the BPS Skyrme model equipped with a partial contribution from the Dirichlet and the Skyrme term. This is the maximal extension of the BPS Skyrme model such that it admits a reduction to an ODE for the profile function $\xi$ with the angular dependence solved by the ansatz $u=v(\theta)e^{iB\varphi}$ and with the same $v(\theta)=\tan \frac{\theta}{2}$.

First of all, let us observe that we can use the topological bounds for both parts of the model separately, i.e. for the second coupled BPS submodel \label{E20-bound} and for the BPS Skyrme model
\be
E_{BPS} \geq \frac{64 \sqrt{2}\pi}{15} |B|.
\ee
Then the improved bound reads 
\be
E^{(2)}+E_{BPS} \geq \left( 4\pi^2 +\frac{64 \sqrt{2}\pi}{15} \lambda m \right) |B| . \label{E2+BPS-bound}
\ee

In Fig. \ref{L2L_BPS-xi} we show the profile function $\xi$ for some particular values of $\lambda$ for charge one solutions. 
For decreasing $\lambda$ the solution approaches the previously found step-function like solution of the $E^{(2)}+E_0$ model. In Fig. \ref{L2L0+L6} we plot the energy per baryon charge  and energy per topological bound (\ref{E2+BPS-bound}) for the first Skyrmions ($B=1,2,3,4$) as a function of the coupling constant $\lambda$. Here we have chosen the mass parameter $m=1.$ As one may expect, the ratio $E/E_{bound}$ tends to 1 as $\lambda \rightarrow 0$. For increasing $\lambda$ the ratio grows as we depart from the BPS theory. Finally, for very large $\lambda$ the ratio  drops again. However, even for extremely large $\lambda$ it is significantly above 1. This means that even in this limit the quadratic term $4\pi \int dr r^2 \xi'^2$ provides an non-negligible contribution to the total energy, and we do not approach the pure BPS regime.

%%%%%%%%%%%%%%%%%%%%%%%%%%%%%%%%%%%%%%%%%
\section{The $\mathcal{L}^{(1)}_{24}$ BPS model and its  non-BPS extension}
%%%%%%%%%%%%%%%%%%%%%%%%%%%%%%%%%%%%%%%%%
Here we consider an extension of the first coupled BPS submodel by the inclusion of a potential 
\be
\mathcal{L}=\mathcal{L}_{24}^{(1)} +m^2 \mathcal{L}_0 .
\ee
For $m=0$, inserting the separation of variable ansatz $\xi (r)$ and $u(\theta,\varphi)$ leads to a complete factorisation of the energy density into an angular part, which is equivalent to the CP(1) model on $\mathbb{S}^2$, and a radial part. For $m\not= 0$ this is no longer true. Instead, the $\mathcal{L}_{24}^{(1)} $ term contains an angular factor proportional to the topological charge density on $\mathbb{S}^2$, whereas the potential part has no angular dependence at all for potentials of the form $\mathcal{U}(\xi)$. 
After the separation of variables, the variation w.r.t. $u$ gives rise
to Euler-Lagrange (EL) equations which can be identified with those of the CP(1)
model and have solutions given by rational maps.
The EL equation for $\xi$, however, is the sum of one angular-dependent term and one angular independent term, which is not compatible with the separation of variables. The only exception is the spherically symmetric charge one case $u=z$, where the topological charge density is a constant. 

Therefore, in this section we will study how the inclusion of the potential influences the properties of the compacton in the $B=1$ sector. The corresponding reduced energy functional is
\bea
E^{(1)}+E_0&=&4\pi \int dr  \left( 2 \sin^2 \xi (1+\xi'^2)+r^2m^2 \mathcal{U}(\xi) \right) \\
&=& 4\pi \int dr  \left( 2 \sin^2 \xi  \xi'^2+2 \sin^2 \xi  + r^2m^2 \mathcal{U}(\xi) \right) \\
&=& 4\pi \int dr  \left( 2 \eta'^2 - 2\eta^2 + 4\eta +r^2m^2 \mathcal{U}(\eta)\right) .
\eea
Adding a potential means adding a new attractive force into the system. Therefore, at least for the case of compact solutions, the size of the compactons will decrease. We find an analytical understanding of this property in the following subsection. 
Furthermore, we can analytically study how the potential breaks the BPS
property.

%%%%%%%%%%%%%%%%%%%%%%%%%%%%%%%%%%%%%%%%%
\subsection{The pion mass potential }
%%%%%%%%%%%%%%%%%%%%%%%%%%%%%%%%%%%%%%%%%
For the usual pion mass potential, $\mathcal{U}_\pi=(1-\cos\xi)=\eta$, the model is 
\be
E\equiv E^{(1)}+E_0 = 4\pi \int dr  \left( 2 \eta'^2 - 2\eta^2 + (4+r^2m^2) \eta \right).
\ee

This energy integral has a unique vacuum at $\eta=0$ in the relevant interval $0\le \eta \le 2$, and the effective potential $V=(4+m^2 r^2)\eta$ approaches its vacuum linearly, that is, $V\sim \eta$ for $\eta \rightarrow 0$. As a result, the static solutions in 
this BPS submodel are compactons.

\begin{figure}
\hspace*{-1.0cm}
\includegraphics[height=5.cm]{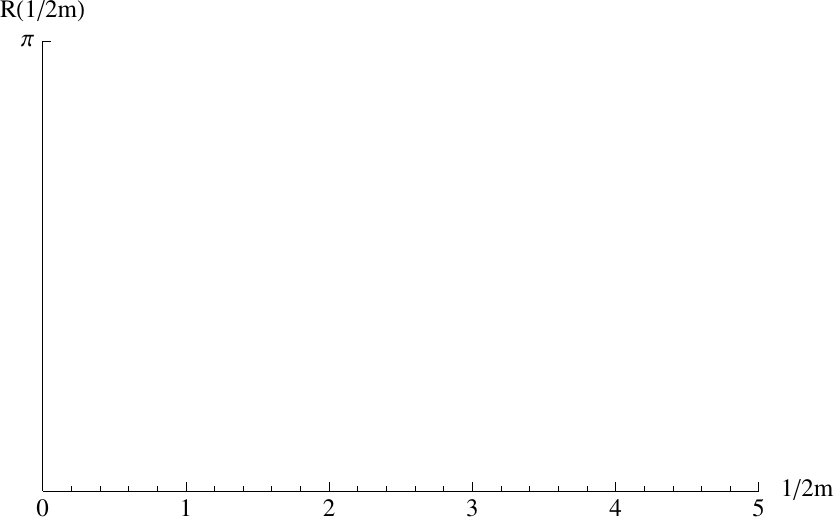}
\includegraphics[height=5cm]{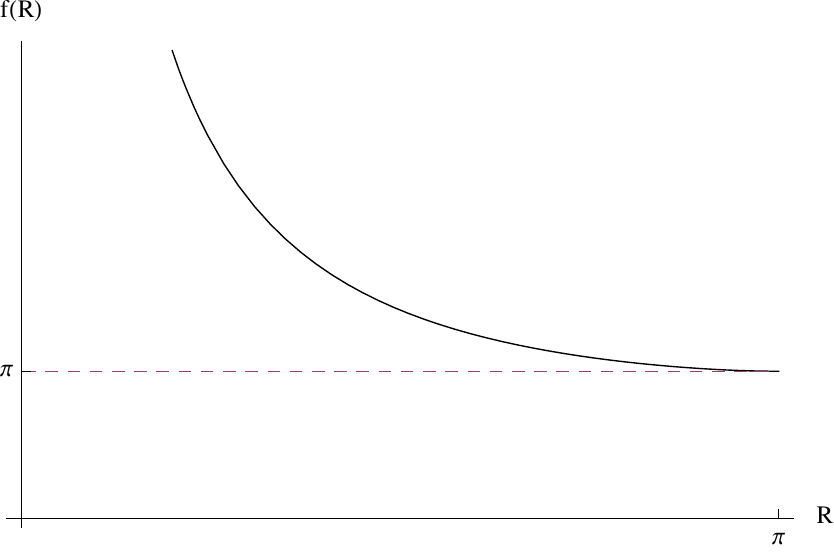}
\caption{Dependence of the radius and $f(R)$ function of the $B=1$ Skyrmion in the $\mathcal{L}^{(1)}_{24}+\mathcal{L}_\pi$ model  on $1/2m$.}
\label{R}
\end{figure}

The corresponding field equation is
\be \label{L1-mass-ELeq}
\eta'' +\eta - \left( 1+ \frac{m^2}{4  } r^2 \right)   =0,
\ee
with the general solution
\be
\eta=\alpha \sin r +\beta \cos r +\frac{m^2}{4}r^2+1-\frac{m^2}{2}.
\ee
Now, we have to impose the proper boundary conditions for compactons (\ref{boundary}). 
Hence,
\bea
\beta&=&1+\frac{m^2}{2}, \\
0&=& \alpha \sin R +\beta \cos R +\frac{m^2R^2}{4}+1-\frac{m^2}{2}, \\
0&=& \alpha \cos R -\beta \sin R +\frac{m^2R}{2}.
\eea
This leads to
\be
\alpha = \left( 1+\frac{m^2}{2}\right) \tan R -\frac{m^2}{2} \frac{R}{\cos R},
\ee
where $R$ is given by the following implicit formula whose graph is shown in Fig. \ref{R}
\be \label{rad}
\frac{m^2}{2}= \frac{1+\cos R}{R\sin R +\cos R -\frac{1}{2}R^2\cos R -1}.
\ee
One can check that $R$ is a monotonously decreasing function of the mass $m$. 
In the limit $m=0$ we find the BPS submodel result $R=\pi$, as expected. Furthermore,
the expansion for small $m$ gives
\be
R=\pi -  \frac{\sqrt{\pi^2-4}}{2}  m + o\left( m\right).
\ee
On the other hand increasing the mass parameter, and thereby increasing the attractive force, shrinks the size 
of the Skyrmion according to
\be
R= 2\sqrt[4]{2} \frac{1}{m^{1/2}} + o\left(\frac{1}{m^{1/2}}\right) \;\;\; \mbox{for} \;\;\; m\rightarrow \infty.
\ee
This solution leads to the following exact expression for the energy
\be
E=8\pi f(R),
\ee
where $f(R)$ is displayed in Fig. \ref{R} and is given by
\bea
f(R)&=&  \frac{1}{30(2+(R^2-2)\cos R - 2R\sin R)^2} \times   \left[ 240R -20R^3+14R^5 - \right.
\nonumber\\ && \nonumber \\
 && \left. 8R^3(R^2-10) \cos R + 
4R(60-35R^2+2R^4)\cos 2R + 40 R^2(R^2- 6) \sin R - \right.
\nonumber \\ && \nonumber \\
&& \left.(240(1-R^2) +55 R^4 )\sin 2R \right], 
\eea
and the size $R$ of the soliton depends on the mass $m$ by \eqref{rad}. This function is a monotonously decreasing function of $R.$  For $R\rightarrow \pi$, which coincides with the limit $m\rightarrow 0$, it tends to its minimal value $f(\pi)=\pi$. This gives $E=8\pi^2 $ for $m=0$. More precisely, for $R\rightarrow \pi$ we find
\be
f(R)= \pi +\frac{\pi}{6} \frac{\pi^2-6}{\pi^2-4} (R-\pi)^2 + o((R-\pi)^2).
\ee
Similarly, for small radius
\be
f(R)=\frac{128}{21} \frac{1}{R} + O(R).
\ee
Combining this with the relation between the size of the compacton and the mass leads to
\be
E(B)=8\pi^2 \left(1+ \frac{\pi^2-6}{6}  m^2  + o\left( m^2 \right)\right), \;\;\; m \rightarrow 0.
\ee
As was observed above, rational maps are not solutions of the problem for $B>1$. This leads to several interesting questions about the case of higher $B$. 
First of all, we may calculate higher $B$ solutions approximately within the rational map ansatz approximation as is done for the standard Skyrme model $\mathcal{L}_{24}$. That is to say, 
we assume separation of variables and also assume that $u$ is given by a rational map.
 Then we integrate over the angular part of the energy functional where the integral over the CP(1) charge density just gives $B$. After the replacement $m^2 \to (m^2/B)$, the resulting reduced energy functional for $\xi (r)$ leads to the same EL equation (\ref{L1-mass-ELeq}) and, therefore, to the same solution and boundary conditions.

There is, however, one important difference between the rational map ansatz in the massless standard Skyrme model and our case. In the standard Skyrme model, the Skyrme term (more precisely, the $\mathcal{L}_{24}^{(2)}$ contribution to it) selects one rational map as its minimiser which determines the optimal rational map and, therefore, the symmetry of the corresponding Skyrmion. In our case, the potential does not depend on $u$ at all and, therefore, cannot lift the degeneracy between arbitrary rational maps within the rational map ansatz approximation. 
This leads to the second question 
about the geometry of the energy minimizers in each topological sector. 
The potential term is not compatible with the separation of variables, so it will most likely lift the rather large degeneracy of the $\mathcal{L}_{24}^{(1)}$ model for true (numerically calculated) Skyrmions. The resulting Skyrmions will, therefore, have definite shapes and symmetries. These symmetries will in general be different from the symmetries of Skyrmions in the ${\cal L}_{24}$ model. The two different symmetries from the mass term in the model considered here, on the one hand, and the massless standard Skyrme model, on the other hand, and their competing effects might be useful to understand the shapes of standard Skyrmions with massive pions. Similar considerations hold for other potential terms. In any case,
this issue requires full three dimensional numerical calculation, which goes beyond the scope of the present paper. We plan to investigate this in the future. 
%%%%%%%%%%%%%%%%%%%%%%%%%%%%%%%%%%%%%%%%%
\section{The Skyrme model and the rational map ansatz}
%%%%%%%%%%%%%%%%%%%%%%%%%%%%%%%%%%%%%%%%%
The fact that arbitrary rational maps (\ref{RM}) are solutions of the first coupled BPS submodel provided an explanation why the rational map ansatz works remarkably well for the massless Skyrme model $\mathcal{L}_{24}$ \cite{newBPS}. Here we further investigate this problem. 

\begin{figure}
\hspace*{-1.0cm}
\includegraphics[height=6.cm]{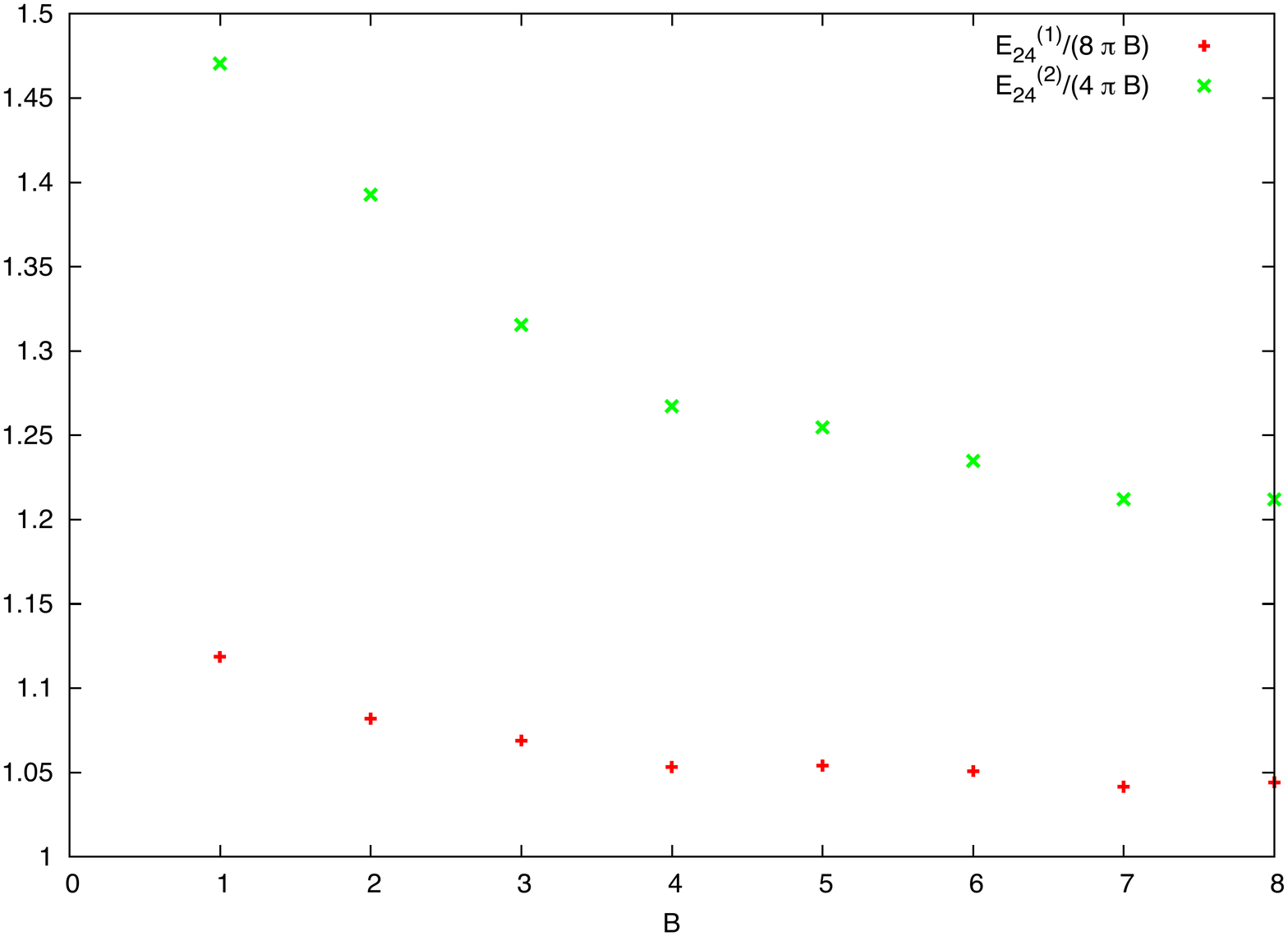}
\includegraphics[height=6.cm]{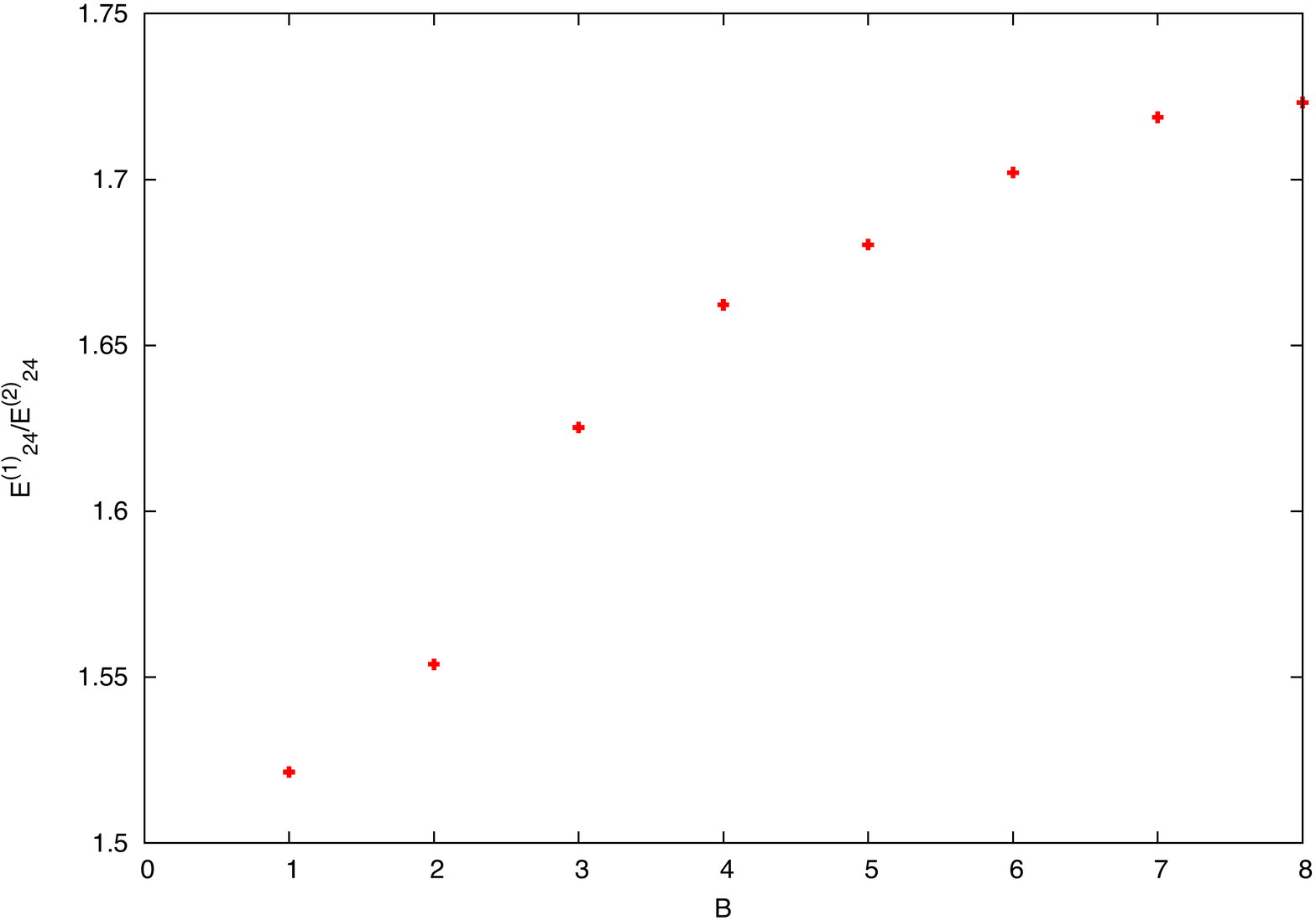}
\caption{Left: On-shell energy of the first and second BPS submodels divided by the corresponding topological bounds as a function of the topological charge. Right: Ratio between the on-shell energies of the first and second BPS submodel.} 
\label{RMA}
\end{figure}

In a first step, we numerically calculate the true soliton solutions of the massless Skyrme model $\mathcal{L}_{24}$ with $B=1, \ldots ,8$. Then we  compute the corresponding on-shell energies of the first $E^{(1)}$ and second $E^{(2)}$ coupled BPS submodels. Obviously, for each $B$ they sum to the energy of the Skyrmion of the massless Skyrme model, $E_{24}=E^{(1)}+E^{(2)}$. It is instructive to plot the on-shell energies divided by the respective topological bounds. Fig. \ref{RMA} shows that the energy of the first BPS submodel computed on the true solution of the full $E_{24}$ model is very close to the topological bound, with the ratio approaching approximately $1.05$ for higher values of the topological charge. This implies that the solution of the full $\mathcal{L}_{24}$ model is rather close to a solution of the $\mathcal{L}^{(1)}$ submodel which can be parametrized by rational maps. On the other hand, the on-shell energy of the second BPS submodel exceeds the corresponding bound much more significantly. This explains why the angular part of a solution of the first BPS Skyrme submodel, rather than the solution emerging from the second BPS submodel, provides a good guess for the true massless Skyrmions. In this approximation, the role of the second submodel $\mathcal{L}^{(2)}$ is to select a minimising rational map among all rational maps of a given degree.  
In addition, Fig. \ref{RMA} shows that the second BPS submodel is the main source of the binding energy problem in the Skyrme model $\mathcal{L}_{24}$. 

The on-shell ratios $E^{(1)}/(8\pi^2 |B|)$ and $E^{(2)}/(4\pi^2 |B|)$ seem to provide a good indication when the rational map ansatz is a good approximation. Indeed, one can compute them for  Skyrme models with arbitrary potentials or with a contribution from the sextic term. If $E^{(1)}/(8\pi^2 |B|)$ is close to one, then the rational map is still a good approximation. It is, however, completely plausible that for many potentials, with or without the inclusion of the sextic term, the ratio is much above one. This should result in a different ansatz for solutions of such a generalized Skyrme model. It is an intriguing question whether in such a situation the second ratio, $E^{(2)}/(4\pi^2 |B|)$, can be made closer to one. Then, the ansatz inherited from the solutions of the second BPS submodel might be the right guess. We emphasize that knowing good analytical approximations for Skyrmions is important for numerical calculations.

\vspace*{0.2cm}

If both bounds were saturated at the same time, then the ratio between the energy of the first and the second BPS submodels would read
\be
\left. \frac{E^{(1)}}{E^{(2)}} \right|_{saturated} = 2.
\ee
As we know, this happens on $\mathbb{S}^3$ base space. For the usual $\mathbb{R}^3$ base space the bounds cannot be simultaneously satisfied. We found that for the true Skyrmions in the massless Skyrme model this ratio grows with the baryon charge but never approaches the ``saturated'' value as shown in Fig. \ref{RMA}.

%%%%%%%%%%%%%%%%%%%%%%%%%%%%%%%%%%%%%%%%%
\section{Conclusions}
%%%%%%%%%%%%%%%%%%%%%%%%%%%%%%%%%%%%%%%%%
In the present paper, we further analysed the coupled BPS submodels of the $\mathcal{L}_{24}$ Skyrme model. 
Although the submodels are coupled, which means that they always co-exist, one can investigate them separately, as some features of Skyrmions in the full theory may originate from one of the submodels. Due to the fact that, by construction, these submodels are examples of BPS theories, they offer unique analytical insights into properties of Skyrmions.
Let us summarise the main results.

\vspace*{0.1cm}

{\it 1. Geometric explanation of the coupled BPS submodels.} A target space coordinate-independent formulation of these models together with the corresponding Bogomolny equations is given by the eigenvalues of the strain tensor. The Bogomomy equations for each of the models are just a subset of the Bogomolny equations emerging in a derivation of the topological Skyrme-Faddeev bound for the Skyrme model. 

\vspace*{0.1cm}

{\it 2. Explanation of the success of the rational map ansatz (RMA).}
The success of the RMA in the construction of approximate solutions of the
standard massless
Skyrme model $\mathcal{L}_{24}$ can be explained by the fact that the on-shell value of the $E^{(1)}$
energy is remarkably close to the relevant topological bound. This implies
that the solution of the full $\mathcal{L}_{24}$ model is rather close to a
solution of the $\mathcal{L}^{(1)}$ submodel which is solved by rational maps.

{\it 3. Thermodynamics at $T=0$.}  We found that, in spite of many differences, both submodels have the same mean-field equation of state (MF EoS) in the high pressure regime $p=\bar{\rho}/3$. As expected, this coincides with the MF EoS for the $\mathcal{L}_{24}$ Skyrme model. Of course, it gives a subleading contribution to the MF EoS of the full Skyrme model $\mathcal{L}_{0246}$ in the high pressure limit, where the thermodynamics is governed by the sextic term \cite{EoS}, \cite{MF}. 
\\
Still, the submodels describe rather different types of Skyrmionic matter. In the first coupled BPS submodel we found a very attractive matter which resulted in compactons  
while in the second submodel the matter reveals a very repulsive nature. In the $\mathcal{L}_{24}$ Skyrme model, where both submodels co-exist, there is a balance between these two opposing properties leading to the appearance of crystal structures \cite{crystal}. Understanding the details of this process requires further analysis. 

\vspace*{0.1cm}

{\it 4. Bose-Einstein condensate.} It is a rather surprising fact that the volume of Skyrmions in the first coupled BPS Skyrme submodel is independent of the topological charge. This behaviour has some similarities with that of a Bose-Einstein condensate. Such a BEC would not be a condensate of individual Skyrmions because it occurs already in the charge one sector. This resembles the perfect fluid property of the BPS Skyrme model where even the $B=1$ solution is a perfect fluid droplet. The observation of a BEC-like sector within the Skyrme model is interesting and definitely deserves further investigation.

\vspace*{0.1cm}

{\it 5. Oscillons.} There is also a rich nontopological sector in the  $\mathcal{L}^{(1)}_{24}$ BPS submodel where approximate exact oscillons 
%and their waying generalisations 
can be found. Notably, they can oscillate for an arbitrarily long time by reducing their amplitude. Therefore, such small amplitude %(energy) 
oscillons dominate interactions in this submodel. Whether they give rise to some nontopological structures in the $\mathcal{L}_{24}$ Skyrme model \cite{zak osc} or can be detected in Skyrmion interactions \cite{Sk scatter} requires further investigations. 

\vspace*{0.1cm}

 {\it 6. Existence of  a solvable non-BPS model.} The second BPS submodel can be extended to various non-BPS theories by the addition of new terms, in such a way that the corresponding ansatz for the $\mathbb{S}^2$ part of the Skyrme field remains valid. As a result, we are left with an ODE for the Skyrme profile function $\xi$ where the topological charge enters as a parameter, which allows for easy studies of such models for any value of the topological charge. 

This is an interesting observation, as it provides an extension of the BPS Skyrme model, where the quadratic as well as quartic terms are partially taken into account, which gives the main contribution to thermodynamical and bulk properties at higher density/pressure \cite{EoS}. This can have obvious applications to nuclear matter, neutron stars  \cite{ns} and hairy black holes \cite{bh}.

\vspace*{0.1cm}

It is widely known that a restriction of the Skyrme model to a two-sphere target space (simply by assuming that $\xi$ takes a constant value which leaves only $u\in \mathbb{C}$ or equivalenty $\bn \in \mathbb{S}^2$ degrees of freedom) gives the Skyrme-Faddeev-Niemi model conjectured to be relevant for the low energy sector of the quantum $SU(2)$ Yang-Mills theory and therefore a candidate for a model of glueballs \cite{niemi}. It is a matter of fact that each of the coupled BPS submodels contains a different contribution from the SFN model. The first BPS submodel contains the $(\partial_\mu \bn)^2$ term which is the kinetic part generated by the dimension two condensate, while the second BPS submodel contains the $(\partial_\mu \bn \times \partial_\nu \bn)^2$ term. This term follows from the YM action if the Faddeev-Niemi-Cho-Shabanov decomposition of the gauge field is assumed \cite{niemi}, \cite{cho}. In particular, it carries magnetic monopole like degrees of freedom of the original gauge field, which coincides with previous remarks that the second BPS submodel resembles a dilaton-magnetic monopole system \cite{jutta}. It would be very desirable to better understand a possible relation between the coupled BPS structures of the Skyrme model and the Skyrme-Faddeev-Niemi action. 

%%%%%%%%%%%%%%%%%%%%%%%%%%%%%%%%%%%%%%%%%
\section*{Acknowledgements}
%%%%%%%%%%%%%%%%%%%%%%%%%%%%%%%%%%%%%%%%%
The authors acknowledge financial support from the Ministry of Education, Culture, and Sports, Spain (Grant No. FPA 2014-58-293-C2-1-P), the Xunta de Galicia (Grant No. INCITE09.296.035PR and Conselleria de Educacion), the Spanish Consolider-Ingenio 2010 Programme CPAN (CSD2007-00042), and FEDER. AW thanks P. Sutcliffe for discussion. SK is grateful to W. Grummitt for helpful suggestions. DF would like to thank the Leverhulme Trust Research Programme Grant RP2013-K-009 SPOCK: Scientific Properties of Complex Knots.

\end{document}